\documentclass[12pt]{article}
\begin{document}
\title{Asymptotic Expansion \\
for the Magnetoconductance Autocorrelation Function}
\author{
\\ \\
Z. Pluha\v r\\
Charles University, Prague, Czech Republic\\ \\
and\\ \\
H.A. Weidenm\"uller\\
Max-Planck-Institut f\"ur Kernphysik, \\
D-69029 Heidelberg, Germany\\ \\ \\}
\date{\today}

\maketitle

\begin{abstract}

We complement a recent calculation (P.B. Gossiaux and the present
authors, Ann. Phys. (N.Y.) in press) of the autocorrelation function
of the conductance versus magnetic field strength for ballistic
electron transport through microstructures with the shape of a
classically chaotic billiard coupled to ideal leads. The function
depends on the total number $M$ of channels and the parameter $t$
which measures the difference in magnetic field strengths. We
determine the leading terms in an asymptotic expansion for large $t$
at fixed $M$, and for large $M$ at fixed $t/M$. We compare our results
and the ones obtained in the previous paper with the squared
Lorentzian suggested by semiclassical theory.

\end{abstract}

\newpage

\section{Introduction}

Lately, the transport of ballistic electrons through microstructures
with the shape of a classically chaotic billiard has received much
attention \cite{ree89,bee91}. One observable investigated both
experimentally and theoretically is the conductance autocorrelation
function $C(\Delta B)$. With $g(B)$ the dimensionless conductance, $B$
the external magnetic field, and $\delta g(B) = g(B) - \overline{g(B)}$
the difference between $g(B)$ and its mean value, $C$ is defined as
$C(\Delta B) = \overline{\delta g(B) \delta g(B+\Delta B)}$. The bars
indicate an average which experimentally is taken over the Fermi
energy or the applied gate voltage and, in theories which simulate the
chaotic billiard in terms of an ensemble of random matrices, over that
ensemble. For ideal coupling between leads and microstructure, $C$
depends only on the number $M$ of channels in the leads, and on $t$, a
measure of $\Delta B$ defined below.

The exact dependence of $C$ on $M$ and $t$ is not known. In the
semiclassical approximation, applicable for $M \gg 1$, $C$ was found
to have the form of a squared Lorentzian \cite{jal90}. This result was
confirmed by calculations using the supersymmetry technique
\cite{efe95,fra95}. However, the supersymmetric nonlinear sigma model
leads to expressions for $C(B)$ which so far have resisted all
attempts at an exact evaluation. In the actual experiments
\cite{Mar92,cha94}, $M$ is quite small. These facts and the desire to
attain a deeper understanding of the supersymmetry technique prompted
us \cite{gos98} to calculate analytically the leading terms in the
asymptotic expansion of $C$ for small $t$ at fixed $M$. In the present
paper, we complement this study by calculating analytically the
leading terms in the asymptotic expansion of $C$ for large $t$ and
fixed $M$, and for large $M$ and fixed $t/M$.

Starting point is the random matrix model for the Hamiltonians of the
chaotic microstructure for two values of the external field $B$. Our
approach is identical to that taken in Ref.~\cite{gos98}. However, we
introduce a new parametrization of the coset manifold suitable for
large values of $t$. We pay particular attention to the Efetov--Wegner
terms (Refs.~\cite{Efe83} to \cite{zir95}) generated by this procedure. 

In order to save space, we keep the introductory part as brief as
possible and refer the reader to our previous paper \cite{gos98}. In
Section \ref{formulation}, we start with the form of the
autocorrelation function obtained after averaging over the ensemble,
after using the saddle--point approximation, and after integration
over the massive modes. Unless otherwise stated, all quantities
appearing in that section are defined as in Ref.~\cite{gos98}. The
parametrization of the coset manifold, and the integral theorem used
in calculating the asymptotic formulae, are presented in Section
\ref{par-of-man}. Simplifications of the integral theorem are
discussed in Section \ref{efe-weg}. The asymptotic formula for large $
t $ at fixed $ M $ is derived in Section \ref{t>>1}. Section
\ref{conclusions} contains the conclusions. Additional mathematical
details including the derivation of the asymptotic formula for large
$M$ at fixed $t/M$ are given in several Appendices.

\section{Supersymmetric Form of the Autocorrelation Function}
\label{formulation}

The two magnetic field strengths are denoted by $B^{(1)}$ and
$B^{(2)}$. We write $g(B^{(i)}) = g^{(i)}$ with $i = 1,2$. The
autocorrelation function 
\begin{equation}
C(t,M)
=\overline{\delta g^{(1)}\delta g^{(2)}}
=\overline{g^{(1)}g^{(2)}}-\overline{g^{(1)}}\,\overline{g^{(2)}}
\label{C-definition}
\end{equation}
depends on $M$, the total number of channels in both leads, and on the
parameter $t$ which is related to the area $A$ of the billiard by
$\sqrt{t} = k |B^{(1)} - B^{(2)}| A / (2 \phi_0)$. Here $k$ is a
numerical constant of order unity and $\phi_0 = hc/e$ is the
elementary flux quantum \cite{Plu95}. For sufficiently large values of
$B^{(1)}$ and $B^{(2)}$, we have $\overline{g^{(1)}} =
\overline{g^{(2)}} = M/2$, while $\overline{g^{(1)} g^{(2)}}$ is given
by \cite{gos98}
\begin{equation}
\label{result}
\overline{g^{(1)} g^{(2)}} = \int {\cal D} \mu(T)
\ \exp \left( - \frac{t}{2} \ < (Q \tau_3)^2 > \right)
\ R(Q) \ \mbox{detg}^{-M} (1+QL).
\end{equation}
Here, $Q = T^{-1}LT$. 
In the sequel, we use brackets $< ... >$ as a short-hand
notation for the graded trace.
The matrices $T$ belong to the coset space
$\mbox{U}(2,2/4) /$ $\mbox{U}(2/2) \times \mbox{U}(2/2)$ where
$\mbox{U}(2/2) \times \mbox{U}(2/2)$ is the subgroup of matrices 
in $\mbox{U}(2,2/4)$ which commute with $L$. The symbol ${\cal D}
\mu(T)$ denotes the invariant integration measure for the coset
space. 
The $8 N$ dimensional graded matrices $L$ and $\tau_3$ are
given by 
\begin{equation}
\label{matrices}
L_{\alpha r p, \alpha' r' p'} = (-)^{(1 + p)} \
\delta_{\alpha \alpha'} \delta_{r r'} \delta_{p p'}, \
(\tau_3)_{\alpha r p, \alpha' r' p'} = (-)^{(1 + r)} \
\delta_{\alpha \alpha'} \delta_{r r'} \delta_{p p'} \ .
\end{equation}
The supersymmetry index $\alpha$ distinguishes ordinary complex
(commuting) integration variables ($\alpha = 0$) and Grassmannn
(anticommuting) variables ($\alpha = 1$). Later, the two values 
0 and 1 of the supersymmetry index $\alpha$ will also be denoted by
$b$ and $f$, respectively. The index $r = 1,2$ refers to the two
fields. The index $p = 1,2$ refers to the retarded and the advanced
propagator, respectively. Finally, $ R( Q ) $ denotes the source term
given by the sum
\begin{equation}
\label{source-terms}
R( Q ) = \sum_{j=1}^{4} N_{j} R_{j}( Q ) \ ,
\end{equation}
where $N_{1} = M^{2}/2, N_{2} = M^{4}/4, N_{3} = N_{4} = M^{3}/4 $,
and where
\begin{eqnarray}
&& R_{1}( Q ) = \
< G I^{(11)} G I^{(22)} >< G I^{(12)} G I^{(21)} > \ ,
\nonumber \\
&& R_{2}( Q ) = \
< G I^{(11)} G I^{(12)} >< G I^{(21)} G I^{(22)} > \ ,
\nonumber \\
&& R_{3}( Q ) = \
< G I^{(11)} G I^{(22)} G I^{(21)} G I^{(12)} > \ ,
\nonumber \\
&& R_{4}( Q ) = \
< G I^{(11)} G I^{(12)} G I^{(21)} G I^{(22)} > \ ,
\label{source-terms-1}
\end{eqnarray}
with
\begin{equation}
\label{auxiliary3}
G( Q ) = ( 1 + Q L )^{-1} \ .
\end{equation}
The matrices $I^{(rp)}$ are given by
\begin{equation}
I_{\alpha'r'p',\alpha''r''p''}^{(rp)}
= (-)^{(1+\alpha')} \delta_{rr'} \delta_{pp'}
\delta_{\alpha'\alpha''}\delta_{r'r''}\delta_{p'p''} \ .
\label{I0-def}
\end{equation}
The graded matrices $Q$ can be parametrized in terms of 32 variables, half of
them commuting, the others, anticommuting. In calculations of the 
average two--point function, one typically deals with a total of 16
integration variables. Except for this increase in the number of variables
and for the form of the source terms (which are, of course, specific to
our problem), the form of our result in Eqs.~(\ref{C-definition})
and (\ref{result}) is quite standard. In spite of this similarity,
the increase in the number of variables renders a full analytical
evaluation of the graded integral~(\ref{result}) very difficult.
As pointed out in the Introduction, our analytical work is restricted to
two asymptotic expansions of this integral. We evaluate the leading
term of the asymptotic expansion of $C(t,M)$ in inverse powers of $t$
and fixed $M$, and the leading term of the asymptotic expansion of
$C(t,M)$ in inverse powers of $M$ and fixed $t/M$. Progress in this
calculation depends crucially on the proper choice of the 32 variables
used in parametrizing the matrices $Q$.

\section{Parametrization of the Saddle--Point \\ Manifold}
\label{par-of-man}

In order to calculate the autocorrelation function for large $t$, we
need a new parametrization of the saddle--point manifold different from
the one used in our previous paper \cite{gos98}. Motivated by the
paper by Altland, Iida and Efetov \cite{alt93}, we write our coset
matrices  $ T $  as products of matrices  $ T_{I} $ obtained by
exponentiating the coset generators anticommuting  with $ \tau_{3} $,
and matrices  $ T_{0} $ obtained by exponentiating the coset
generators commuting with $ \tau_{3} $,
\begin{equation} 
T = T_{I} T_{0}  \;,
\label{T-TIT0}
\end{equation}
where  $ T_{0} $  and  $ T_{I} $  are the matrices
\begin{equation}
T_{0}
= \left(
  \begin{array}{cccc}
  t^{11}_{1}   & t^{12}_{1}   & 0   & 0   \\
  t^{21}_{1}   & t^{22}_{1}   & 0   & 0   \\
  0   & 0   & t^{11}_{2}   & t^{12}_{2}   \\
  0   & 0   & t^{21}_{2}   & t^{22}_{2}
  \end{array}
  \right)  \;,
\qquad
T_{I}
= \left(
  \begin{array}{cccc}
  t^{11}_{3}   & 0   & 0   & t^{12}_{3}   \\
  0   & t^{22}_{4}   & t^{21}_{4}   & 0   \\
  0   & t^{12}_{4}   & t^{11}_{4}   & 0   \\
  t^{21}_{3}   & 0   & 0   & t^{22}_{3}
  \end{array}
  \right)  \;,
\label{T0TI-t}
\end{equation}
with
\begin{equation}
t^{21}_{q} = k ( t^{12}_{q} )^{\dagger}  \;,
\qquad
t^{11}_{q} = ( 1 + t^{12}_{q} t^{21}_{q} )^{1/2}  \;, \qquad
t^{22}_{q} = ( 1 + t^{21}_{q} t^{12}_{q} )^{1/2}  \;
\label{t-rel}
\end{equation}
for $q = 1,2,3,4$ and
\begin{equation}
k = \mbox{diag}( 1, -1 )  \;.
\label{k-q}
\end{equation}
In writing the matrices we label the rows of matrices of dimensions 2,
4, and 8 by the indices $\alpha$, $( p \alpha)$ and $ ( r p \alpha )
$, respectively. The indices follow in lexicographical order. The
matrices of dimensions 4 and 8 are presented in block form. By
construction, the matrix  $ T_{0} $ ($ T_{I} $) does (does not)
commute with $ \tau_{3} $. Eq.~(\ref{T0TI-t}) shows that the matrices 
$ T_{0} $ and $ T_{I} $ are constructed from the GUE coset matrices
\begin{equation}
T_{q}
= \left(
  \begin{array}{cc}
  t^{11}_{q}   & t^{12}_{q}   \\
  t^{21}_{q}   & t^{22}_{q}
  \end{array}
  \right) \ ,
\label{Tq-t}
\end{equation}
the matrix $ T_{0} $ from the matrices $ T_{1} $ and $ T_{2} $, the
matrix $ T_{I} $ from $ T_{3} $ and $ T_{4} $. The matrix elements $
(t^{12}_{q})_{\alpha \alpha'} $ and their conjugates $
(t^{12}_{q})_{\alpha \alpha'}^{\ast} $ 
represent the Cartesian coordinates of $ T_{q} $.
We denote this set of coordinates by $x_{q}$.
The invariant measure for integration over the coset matrices
$ T = T_{I} T_{0} $ has the form
\begin{equation}
{\cal D} \mu( T )
= \prod_{q} {\cal D} \mu_{ \mbox {\scriptsize G} }( T_{q} )
\ {\cal P}( T_{3}, T_{4} )  \;,
\label{muT}
\end{equation}
where  $ {\cal D} \mu_{\mbox {\scriptsize G} } ( T_{q} ) $
is the invariant measure for
integration over the GUE coset matrices  $ T_{q} $,  and
$ {\cal P} ( T_{3}, T_{4} ) $  is a function of eigenvalues
of $ t_{3}^{12}t_{3}^{21} $  and $ t_{4}^{12}t_{4}^{21} $  given
later. In Cartesian coordinates, 
\begin{equation}
{\cal D} \mu_{ \mbox {\scriptsize G} } ( T_{q}( x_{q} ) )
= \prod_{\alpha\alpha'}
  \mbox{d}( t^{12}_{q} )_{\alpha \alpha'}
  \mbox{d}( t^{12}_{q} )_{\alpha \alpha'}^{\ast}  \;.
\label{muTq-xq}
\end{equation}
In spite of the simplicity of this expression, the Cartesian
coordinates are not well suited for our calculation. Therefore, we
follow Efetov \cite{Efe83} and change to polar coordinates.

\subsection{Polar Coordinates}
\label{pol-coo}

We introduce polar coordinates $ z_{r} = (\theta_{r}^{b}, \phi_{r}^{b}, 
\theta_{r}^{f}, \phi_{r}^{f}, \gamma_{r}^{1}, \gamma_{r}^{1\ast}, 
\gamma_{r}^{2}, \gamma_{r}^{2\ast} ) $ for the matrix $ T_{0} $ by the
transformation 
\begin{equation}
t^{12}_{r} = u^{1}_{r} \lambda^{12}_{r} ( u^{2}_{r} )^{-1}  \;,
\qquad
t^{21}_{r} = u^{2}_{r} \lambda^{21}_{r} ( u^{1}_{r} )^{-1}  \;
\label{tr-ular}
\end{equation}
where $ u_{r}^{p}, \lambda_{r}^{12}, \lambda_{r}^{21}$ are defined by
\begin{equation}
u^{1}_{r}
= \exp \left(\!
  \begin{array}{cc}
  0   & \gamma^{1}_{r}   \\
  \gamma^{1\ast}_{r}   & 0
  \end{array}
  \!\right)  \;,
\qquad
u^{2}_{r}
= \exp \left(\!
  \begin{array}{cc}
  0   & \gamma^{2\ast}_{r}   \\
  \gamma^{2}_{r}   & 0
  \end{array}
  \!\right)  \;,
\label{ur-gar}
\end{equation}
\begin{equation}
\lambda_{r}^{12} = i \sin (\theta_{r}/2) \mbox{e}^{i\phi_{r}}  \;,  \qquad
\lambda_{r}^{21} = i \sin (\theta_{r}/2) \mbox{e}^{-i\phi_{r}}  \;,
\label{lar-thr-phr}
\end{equation}
\begin{equation}
\theta_{r} = \mbox{diag}( \theta_{r}^{b}, \theta_{r}^{f} )  \;,
\qquad
\phi_{r} = \mbox{diag}( \phi_{r}^{b}, \phi_{r}^{f} )  \;.
\label{thr-phr}
\end{equation}
\\[1mm]
The coordinates  $ \theta_{r}^{\alpha}, \phi_{r}^{\alpha} $ are
commuting variables, the cordinates $ \gamma_{r}^{p},
\gamma_{r}^{p\ast} $ anticommuting variables. The relation $
t_{r}^{21} = k( t_{r}^{12} )^{\dagger} $ implies $ \lambda_{r}^{21} =
k( \lambda_{r}^{12} )^{\ast} $ so that $ \theta_{r} = - k
\theta_{r}^{\ast} $ whereas $ \phi_{r} = \phi_{r}^{\ast} $. 
Substituting Eqs.~(\ref{tr-ular}) into 
Eqs.~(\ref{T0TI-t}), (\ref{t-rel}) yields 
\begin{equation}
T_{0} = U_{0} \Lambda_{0} U_{0}^{-1}
\label{T0-ULa0}
\end{equation}
where 
\begin{equation}
U_{0} = \mbox{diag} ( u^{1}_{1}, u^{2}_{1}, u^{1}_{2}, u^{2}_{2} ) \ ,
\qquad 
\Lambda_{0}
= \left(
  \begin{array}{cccc}
  \lambda^{11}_{1}   & \lambda^{12}_{1}   & 0   & 0   \\
  \lambda^{21}_{1}   & \lambda^{22}_{1}   & 0   & 0   \\
  0   & 0   & \lambda^{11}_{2}   & \lambda^{12}_{2}   \\
  0   & 0   & \lambda^{21}_{2}   & \lambda^{22}_{2}
  \end{array}
  \right) \ ,
\label{ULa0-ula}
\end{equation}
with 
$ \lambda_{r}^{11} = \lambda_{r}^{22} = \cos(\theta_{r}/2) $.
We insert this expression for $ T_{0} $ in  $ T = T_{I} T_{0} $ and get
\begin{equation}
T = U_{0} \tilde T_{I} \Lambda_{0} U_{0}^{-1}\;,
\label{T-ULa0tiTI}
\end{equation}
where
\begin{equation}
\tilde T_{I} = U_{0}^{-1} T_{I} U_{0} \;
\label{tiTI-U0TI}
\end{equation}
has the same form as the matrix $ T_{I} $ but with  $ t_{3}^{pp'},
t_{4}^{pp'} $ replaced by
\begin{equation}
\tilde t_{3}^{pp'} = ( u_{p}^{p} )^{-1} t_{3}^{pp'} u_{p'}^{p'} \ ,
\qquad
\tilde t_{4}^{pp'} = ( u_{\hat p}^{p} )^{-1} t_{4}^{pp'} u_{\hat p'}^{p'} \ .
\label{tits-uts}
\end{equation}
The indices $ \hat i $ are defined by $ \hat i = 2,1 $ for $ i = 1,2 $
respectively. We replace the Cartesian coordinates  $ x_{s} $
of $ T_{s} $ by the Cartesian coordinates  $ \tilde x_{s} $  of
$ \tilde T_{s} $. The Berezinian of this coordinate transformation is
equal to one. Since for $s = 3,4$ the matrices $ \tilde t_{s}^{12}
\tilde t_{s}^{21} $ and $ t_{s}^{12} t_{s}^{21} $ have the same
eigenvalues, the measure has the same form in the old and in the new
coordinates. We suppress the tildes. Finally we pass from the
Cartesian coordinates $x_{s}$ of $T_{I}$ to the polar coordinates
$ z_{s} = ( \theta_{s}^{b}, \phi_{s}^{b}, \theta_{s}^{f}, \phi_{s}^{f},
\gamma_{s}^{1}, \gamma_{s}^{1\ast}, \gamma_{s}^{2}, \gamma_{s}^{2\ast}
) $ . We do this in the same way as in the transformation from $ x_{r}
$  to $ z_{r} $. Thus we get
\begin{equation}
T_{I} = U_{I} \Lambda_{I} U_{I}^{-1} \;,
\label{TI-ULaI}
\end{equation}
where $ U_{I}, \Lambda_{I} $ denote the counterparts of
$ U_{0}, \Lambda_{0} $ defined by
\\[1mm]
\begin{equation}
U_{I} = \mbox{diag} ( u^{1}_{3}, u^{2}_{4}, u^{1}_{4}, u^{2}_{3} ) \ ,
\qquad
\Lambda_{I}
= \left(
  \begin{array}{cccc}
  \lambda^{11}_{3}   & 0   & 0   & \lambda^{12}_{3}   \\
  0   & \lambda^{22}_{4}   & \lambda^{21}_{4}   & 0   \\
  0   & \lambda^{12}_{4}   & \lambda^{11}_{4}   & 0   \\
  \lambda^{21}_{3}   & 0   & 0   & \lambda^{22}_{3}
  \end{array}
  \right) \ ,
\label{ULaI-ula}
\end{equation}
with
$ \lambda^{11}_{s} = \lambda^{22}_{s} = \cos (\theta_{s}/2) $.
Collecting the results yields the matrix $ T $ as function
$ T = T( z ) $  of the polar coordinates
$ z = (z_{1}, z_{2}, z_{3}, z_{4}) $,
\begin{equation}
T = U_{0} U_{I} \Lambda_{I} U_{I}^{-1} \Lambda_{0} U_{0}^{-1} \ .
\label{T-ULa}
\end{equation}
For all $q$, the matrices $ u_{q}^{p}, \lambda_{q}^{12} $ and $
\lambda_{q}^{21} $ have the form shown in Eqs.~(\ref{ur-gar}) and
(\ref{lar-thr-phr}). By construction, the matrices $ U_{0},
\Lambda_{0} $ depend solely on the coordinates $ z_{12} = ( z_{1},
z_{2} )$, the matrices $ U_{I}, \Lambda_{I} $ solely on the
coordinates $ z_{34} = ( z_{3}, z_{4} )$. Moreover, $ U_{0}, U_{I}
$ depend only on the anticommuting variables, $ \Lambda_{0},
\Lambda_{I} $ only on the commuting variables. Substituting
Eq.~(\ref{T-ULa}) into $ Q = T^{-1} L T $ and making use of the
commutativity of $ U_{0}, U_{I} $ with  $ L $ yields
\begin{equation}
Q = U_{0} \Lambda_{0}^{-1} U_{I} \Lambda_{I}^{-1}
    L \Lambda_{I} U_{I}^{-1} \Lambda_{0} U_{0}^{-1}   \;.
\label{Q-ULa}
\end{equation}
This is the expression of the matrix  $ Q $  in polar coordinates.

The integration measure $\mbox{d}\mu(z) = {\cal D} \mu( T( z ) ) $
has the form
\begin{equation}
\mbox{d}\mu( z ) 
= \prod \nolimits_{q} \mbox{d}\mu_{\mbox{\scriptsize G}}( z_{q} )
  \ \rho( \theta_{34} )
\label{muz}
\end{equation}
where the $ \mbox{d} \mu_{ \mbox{\scriptsize G}} ( z_{q} )
= {\cal D} \mu_{ \mbox{\scriptsize G}} ( T_{q}( z_{q} ) ) $
denote the GUE measures given by
\begin{equation}
\mbox{d}\mu_{ \mbox{\scriptsize G} }( z_{q} ) =
\mbox{d}[ z_{q} ] \rho_{ \mbox{\scriptsize G} }( \theta_{q} )
\label{muTq-zq}
\end{equation}
with
\begin{eqnarray} &&
\mbox{d} [ z_{q} ] = \mbox{d}[ \theta_{q} ] \mbox{d} [ \chi_{q} ] \ ,
\qquad
\rho_{ \mbox{\scriptsize G} }( \theta_{q} ) = \prod \nolimits_{\alpha}
\sin \theta_{q}^{\alpha} \ < \cos \theta_{q} >^{-2} \ ,
\label{dzq-rhG}
\\ &&
\qquad \mbox{d}[ \theta_{q} ]
= \prod \nolimits_{\alpha} \mbox{d}\theta_{q}^{\alpha} \ , \qquad
\mbox{d} [ \chi_{q} ] 
=\prod \nolimits_{\alpha} \mbox{d}\phi_{q}^{\alpha} \, 
\prod \nolimits_{p} \mbox{d}\gamma_{q}^{p} \mbox{d}\gamma_{q}^{p\ast} \ ,
\label{dthq-dchq}
\end{eqnarray}
and where
$ \rho( \theta_{34} ) = {\cal P}( T_{3}(z_{3}), T_{4}(z_{4}) ) $
denotes the density
\begin{equation}
\rho( \theta_{34} )
=\prod \nolimits_{\alpha} 
( \cos\theta_{3}^{\alpha} + \cos\theta_{4}^{\alpha} )^{2} \
\prod \nolimits_{\alpha \neq \alpha'}
( \cos\theta_{3}^{\alpha} + \cos\theta_{4}^{\alpha'} )^{-2} \ . \qquad
\label{rh-th34}
\end{equation}
>From these equations, 
\begin{equation}
\mbox{d} \mu( z )
= \mbox{d}[ z ] \prod \nolimits_{q} 
\rho_{ \mbox{\scriptsize G} }( \theta_{q} ) \ \rho( \theta_{34} ) \ ,
\qquad
\mbox{d}[z] = \prod \nolimits_{q} \mbox{d}[ z_{q} ] \ .
\label{muz-dz}
\end{equation}
The ordinary parts of the angles $\theta_{q}^{b}$ are integrated over
the positive imaginary axis, the ordinary parts of the phases
$\phi_{q}^{\alpha}$ over an interval of length $2\pi$. The ordinary
parts of $\theta_{r}^{f}$ are integrated over the interval $(0,\pi)$,
those of $\theta_{s}^{f}$ over the interval $(0,\pi/2)$.  

An essential simplification of the calculation results from the
observation that the matrices $ T $ possess a high degree of symmetry.
We consider similarity transformations induced by the idempotent
permutation matrices 
\begin{equation}
J_{12}
= \left(
  \begin{array}{cccc}
  0   & 0   & 0   & 1   \\
  0   & 0   & 1   & 0   \\
  0   & 1   & 0   & 0   \\
  1   & 0   & 0   & 0
  \end{array}
  \right)  \;,
\qquad
J_{34}
= \left(
  \begin{array}{cccc}
  0   & 1   & 0   & 0   \\
  1   & 0   & 0   & 0   \\
  0   & 0   & 0   & 1   \\
  0   & 0   & 1   & 0
  \end{array}
  \right)  \;.
\label{J12-J34}
\end{equation}
It is easy to see that $U_{0}, \Lambda_{0}, U_{I}$ and $\Lambda_{I}$
satisfy the symmetry relations
\begin{eqnarray} &&
J_{12} U_{0}( z_{12} ) J_{12} = U_{0}( w_{21} ) \ , \qquad
J_{12} \Lambda_{0}( z_{12} ) J_{12} = \Lambda_{0}( w_{21} ) \ ,
\nonumber \\ &&
J_{12} U_{I}( z_{34} ) J_{12} = U_{I}( w_{34} ) \ , \qquad
J_{12} \Lambda_{I}( z_{34} ) J_{12} = \Lambda_{I}( w_{34} ) \ ,
\label{ULa-J12}
\end{eqnarray}
\begin{eqnarray} &&
J_{34} U_{0}( z_{12} ) J_{34} = U_{0}( w_{12} ) \ , \qquad
J_{34} \Lambda_{0}( z_{12} ) J_{34} = \Lambda_{0}( w_{12} ) \ ,
\nonumber \\ &&
J_{34} U_{I}( z_{34} ) J_{34} = U_{I}( w_{43} ) \ , \qquad
J_{34} \Lambda_{I}( z_{34} ) J_{34} = \Lambda_{I}( w_{43} ) \ ,
\label{ULa-J34}
\end{eqnarray}
where 
$ w_{q}
= ( \theta_{q}^{b}, \psi_{q}^{b},
    \theta_{q}^{f}, \psi_{q}^{f},
    \beta_{q}^{1}, \bar \beta_{q}^{1},
    \beta_{q}^{2}, \bar \beta_{q}^{2} ) $
represents a second set of polar coordinates defined by 
\begin{equation}
\psi_{q}^{\alpha} = - \phi_{q}^{\alpha} \;,  \qquad
\beta^{p}_{q} = \gamma^{\hat p\ast}_{q}  \;,  \qquad
\bar \beta^{p}_{q} = \gamma^{\hat p}_{q}  \;.
\label{psbe-phga}
\end{equation}
For later developments it is important to consider the coordinates
$ w_{q} $ on the same footing as the coordinates $ z_{q} $. The
Berezinian of the coordinate transformation from $ z $ to $ w $ is
equal to unity, and the coordinates $w$ are integrated with the same
measure over the same domain as the cordinates $z$. Inserting the
symmetry relations~(\ref{ULa-J12}) and (\ref{ULa-J34}) into
Eq.~(\ref{T-ULa}) yields the symmetry relations
\begin{equation}
T( z_{12}, z_{34} ) =
J_{12} T( w_{21}, w_{34} ) J_{12}
= J_{34} T( w_{12}, w_{43} ) J_{34} \ .
\label{Tz-J12J34}
\end{equation}
Since the permutation matrices  $ J_{12}, J_{34} $  anticommute with
the matrix $ L $, the corresponding symmetry relations for the
matrices $Q=T^{-1}LT$ read
\begin{equation}
Q( z_{12}, z_{34} ) =
- J_{12} Q( w_{21}, w_{34} ) J_{12}
= - J_{34} Q( w_{12}, w_{43} ) J_{34} \ .
\label{Qz-J12J34}
\end{equation}

\subsection{Integral Theorem}
\label{int-the}

Eqs.~(\ref{muTq-zq}) and (\ref{dzq-rhG}) show that the measures
$ \mbox{d}\mu_{\mbox{\scriptsize G}}( z_{q} ) $ contain
nonintegrable singularities located at $ < \cos \theta_{q} >
= 2(| \lambda_{q}^{b} |^{2} + | \lambda_{q}^{f} |^{2}) = 2 < t^{12}_{q}
t^{21}_{q} > = 0 $.
As we evaluate the
integral
\begin{equation}
{\cal I}[ F ] = \int {\cal D} \mu( T ) F( Q )
\label{I-int}
\end{equation}
for some function $ F( Q ) $ over the coset space in polar
coordinates, the singularities cause the occurrence of additional
terms (the Efetov--Wegner terms). These terms can be found by applying
the method used in Ref.~\cite{gos98}. We write the integral in
Cartesian coordinates and exclude an infinitesimal neighbourhood $
|(t_{q}^{12})_{11}|^{2} + |(t_{q}^{12})_{22}|^{2} < \varepsilon $ of
singularities of $ \mbox{d}\mu_{\mbox{\scriptsize G}}( z_{q} ) $ from
the domain of integration. Making use of Berezin's theory \cite{ber87}
of coordinate transformations in superintegrals over functions which
do not vanish on the boundary of the integration region, we change to
polar coordinates, let $ \varepsilon $ go to $0^+$, and obtain
\begin{equation}
{\cal I}[ F ]
= \int \mbox{d}[ z ] \rho( \theta_{34} )
  \prod \nolimits_{q} \Big( \delta( z_{q} )
   + \rho_{ \mbox{\scriptsize G} }( \theta_{q} ) P( z_{q} ) \Big)
  \ F( Q ) \ .
\label{IF-pol}
\end{equation}
Here $ \delta ( z_{q} ) $ denotes the $ \delta $ function of the polar
coordinate $ z_{q} $, and $ P( z_{q}) $ is the projector on the subspace
of functions of 4th order in the anticommuting coordinates
$\gamma^{1}_{q}, \gamma^{1\ast}_{q}, \gamma^{2}_{q} $ and 
$\gamma^{2\ast}_{q} $. For a function $f(z_{q})$ of $z_{q}$ which is 
regular at $ z_{q} = 0 $ and whose Taylor expansion in the
anticommuting variables terminates with the 4th order term
$f_{4}(z_{q})$, we have
\begin{equation}
\int \mbox{d} [ z_{q} ] \delta ( z_{q} )\ f( z_{q} )
     = f( 0 )  \;,
\quad
P( z_{q} )\ f( z_{q} ) = f_{4}(z_{q}) \;.
\label{dezq-Pzq}
\end{equation}
The product over the four factors $ \Big( \delta( z_{q} ) +
  \rho_{ \mbox{\scriptsize G} }( \theta_{q} ) P( z_{q} ) \Big) $
in the integral~(\ref{IF-pol}) yields a sum of sixteen terms. We
specify these terms by the indices $ a = (a_{1} a_{2} a_{3} a_{4})$.
We put $ a_{q} = 0 $ ($ a_{q} = 1 $) if the term contains 
$\delta(z_{q} )$ ($ \rho_{ \mbox{\scriptsize G} }( \theta_{q} )
P( z_{q} )$, respectively). This yields
\begin{equation}
{\cal I}[ F ]
= \sum_{ a_{1}, a_{2}, a_{3}, a_{4} = 0,1 }
  {\cal I}_{ a_{1} a_{2} a_{3} a_{4} }[ F ]  \;
\label{IF-IaF}
\end{equation}
where
\begin{eqnarray} &&
{\cal I}_{ a_{1} a_{2} a_{3} a_{4} }[ F ]
\nonumber \\ &&
= \int \mbox{d}[ z ]
\rho_{ a_{1} a_{2} a_{3} a_{4} }( \theta_{12}, \theta_{34} )
\prod \nolimits_{q}^{(0)} \delta( z_{q} )
\prod \nolimits^{(1)}_{q} P( z_{q} ) \ F( Q ) 
\label{IaF}
\end{eqnarray}
with
\begin{equation}
\rho_{ a_{1} a_{2} a_{3} a_{4} }( \theta_{12}, \theta_{34} ) =
\prod \nolimits^{(1)}_{q} \rho_{ \mbox{\scriptsize G} }( \theta_{q} )
\rho( \theta_{34} ) \ .
\label{rha}
\end{equation}
The product denoted by $\prod^{(0)}_{q}$ ($\prod^{(1)}_{q}$) extends
over those values of $q$ for which $a_q =0$ ($a_q =1$, respectively)
and equals unity if none of the $a_q$'s meets the definition. 
Eq.~(\ref{IF-IaF}) is the desired decomposition of the integral
(\ref{I-int}) into surface and volume terms. The integral $ {\cal
  I}_{1111}[ F ]$ extends over the entire coset space and represents
the volume term. The remaining fifteen integrals represent the
boundary or Efetov--Wegner terms. The term $ {\cal I}_{0000}[F]$ is
equal to the value of $ F( Q( z ) ) $ at $ z = 0 $, $ {\cal I}_{0000}[
F ] = F( L ) $. For brevity, we use the symbol $ {\cal I}_{a} $ for
the integrals ${\cal I}_{ a_{1} a_{2} a_{3} a_{4} }$.

\section{The Efetov--Wegner Terms}
\label{efe-weg}

In our case, the integrand $ F $ in the integral $ {\cal I}[F] $ has
the form given in Eqs.~(\ref{result}) to (\ref{auxiliary3}). We write
$F$ as the product
\begin{equation}
F = K \,R \,D \ ,
\label{F-KRD}
\end{equation}
where $ K $ denotes the coupling term,
\begin{equation}
K = \mbox{e}^{ -(t/2) < (Q \tau_{3})^{2} > } \ ,
\label{K-Q}
\end{equation}
$ R $  denotes the source term (\ref{source-terms}),
and $ D $  denotes the function
\begin{equation}
D = \mbox{detg}^{-M}( 1 + Q L )
= \mbox{e}^{ -M < \ln ( 1 + Q L ) > } \ .
\label{D-Q}
\end{equation}
We show that for this form of $F$ only four of the integrals ${\cal
  I}_{a}[F]$ give a nonvanishing contribution. 

\subsection{The Integrals ${\cal I}_{a}[F]$}
\label{int-IaF}

We rewrite Eq.~(\ref{Q-ULa}) for $Q$ in the form
\begin{equation}
Q = U_{0} \Lambda_{0}^{-1} Q_{I} \Lambda_{0} U_{0}^{-1} \;, \qquad
Q_{I} = U_{I} \Lambda_{I}^{-1} L \Lambda_{I} U_{I}^{-1} \;
\label{QQI-ULa}
\end{equation}
and use this form to rewrite the terms $K,R$, and $D$. For the
coupling term $ K $, we use that $ U_{0}, \Lambda_{0} $ and $ U_{I} $
commute with $ \tau_{3} $. As a consequence, we have $ < ( Q \tau_{3}
)^{2} > = < ( Q_{I} \tau_{3} )^{2} > = < ( \Lambda_{I}^{-1} L \Lambda_{I}
\tau_{3} )^{2} > $. Using Eqs.~(\ref{ULaI-ula}) and (\ref{lar-thr-phr}), 
we find 
\begin{equation}
K = \mbox{e}^{-\frac{t}{2} < (Q_{I}\tau_{3})^{2} > }
= \mbox{e}^{ 2t \sum_{s} < \sin^{2} \theta_{s} > } \ .
\label{K-ths}
\end{equation}
The exponent of the coupling term $K$ thus depends only on the
angles $\theta_{s}$. The functions $R$ and $D$ both depend on $Q$ via
the matrix $ (1+QL)^{-1} $. We write $ Q_{I} = L + \delta Q_{I} $ and
use Eq.~(\ref{QQI-ULa}). We note that $U_{0}$ commutes with $L$,
introduce the matrix $ G_{0} = ( 1 + \Lambda_{0}^{-1} L \Lambda_{0} L
)^{-1} $, note that $G_{0}$ commutes with $\Lambda_{0}^{-1}$ and that
$G_{0}\Lambda_{0}^{-1}$ is diagonal and therefore commutes with $ L $. 
We find
\begin{equation}
( 1 + Q L )^{-1} =
U_{0} \Lambda_{0}^{-1} G_{0}( 1 + \delta W )^{-1}\ \Lambda_{0} U_{0}^{-1}
\label{(1+QL)inv}
\end{equation}
where
\begin{equation}
\delta W = \delta Q_{I} G_{0} L \, \quad
\label{deW}
\end{equation}
with
\begin{equation}
\delta Q_{I} = Q_{I} - L \ ,
\qquad
G_{0} = ( 1 + \Lambda_{0}^{-1} L \Lambda_{0} L )^{-1} \ .
\label{deQI-G0}
\end{equation}
Inserting the expression for $ ( 1 + Q L )^{-1} $ in the defining
Equation~(\ref{source-terms}) for $ R $ we find that the
anticommuting variables $ \gamma_{r}^{p}, \gamma_{r}^{p\ast} $ appear
only via the matrices 
\begin{equation}
I_{0}^{(rp)} = U_{0}^{-1} I^{(rp)} U_{0} \ .
\label{I0rp}
\end{equation}
The explicit form of these matrices as functions of $\gamma_{r}^{p},
\gamma_{r}^{p\ast}$ can be found with the help of
Eqs.~(\ref{ur-gar}) and (\ref{I0-def}).
We obtain
\begin{equation}
\prod \nolimits^{(0)}_{q} \delta( z_{q} )
\prod \nolimits^{(1)}_{q} P( z_{q} ) \ R D = 
\prod \nolimits^{(0)}_{q} \delta( z_{q} )
\prod \nolimits^{(1)}_{q} P( z_{q} ) \ S D \ ,
\label{PRD-PSD}
\end{equation}
where
\begin{equation}
S = \sum_{j} N_{j} S_{j} \ 
\label{S-NSj}
\end{equation}
and
\begin{eqnarray} &&
S_{1} =
\ < V I_{0}^{(11)} V I_{0}^{(22)} > \ < V I_{0}^{(12)} V I_{0}^{(21)} > \ ,
\nonumber \\ &&
S_{2} =
\ < V I_{0}^{(11)} V I_{0}^{(12)} > \ < V I_{0}^{(21)} V I_{0}^{(22)} > \ ,
\nonumber \\ &&
S_{3} =
\ < V I_{0}^{(11)} V I_{0}^{(22)} V I_{0}^{(21)} V I_{0}^{(12)} > \ ,
\nonumber \\ &&
S_{4} =
\ < V I_{0}^{(11)} V I_{0}^{(12)} V I_{0}^{(21)} V I_{0}^{(22)} > \ ,
\label{Sj-VI0}
\end{eqnarray}
with
\begin{equation}
V = ( 1 + \delta W )^{-1} G_{0} \ .
\label{V-deWG0}
\end{equation}
The anticommuting variables 
$ \gamma_{s}^{p}, \gamma_{s}^{p\ast} $ appear
only in the matrix $ \delta W $. For $D$, we use Eqs.~(\ref{D-Q}),
(\ref{(1+QL)inv}), (\ref{V-deWG0}), (\ref{ULa0-ula}) and
(\ref{lar-thr-phr}), and obtain 
\begin{equation}
D = D_{0}\ \mbox{e}^{ - M < \ln(1 + \delta W) > } \ ,
\label{D-D0deW}
\end{equation}
where 
\begin{equation}
D_{0} = \mbox{e}^{ - M < \ln( 1 + \Lambda_{0}^{-1} L \Lambda_{0} L ) > }
= \mbox{e}^{ - M \sum_{r}< \ln( 1 + \cos\theta_r ) > }
\label{D0-thr}
\end{equation}
depends only on $\theta_{r}$. Again, the anticommuting variables
appear only in $ \delta W$. For the integral ${\cal I}_{a}[ F ]$, we have  
\begin{equation}
{\cal I}_{a}[ F ] 
= \int \mbox{d}[z] \rho_{a} \prod \nolimits_{q}^{(0)} \delta ( z_{q} ) \
\prod \nolimits_{q}^{(1)} P( z_{q} )\, K S D \ ,
\label{IaF-PKSD}
\end{equation}
with $K$, $S$ and $D$ given by Eqs.~(\ref{K-ths}), (\ref{S-NSj}) and 
(\ref{D-D0deW}).
 
For later use we also give explicit expressions for $ \delta Q_{I} $
and $ G_{0} $ in terms of $\theta_{q},\phi_{q}$ and $u_{s}^{p}$.
Using the formulae for $U_{0},\Lambda_{0},U_{I},\Lambda_{I}$ in
Eqs.~(\ref{ULa0-ula}), (\ref{ULaI-ula}) and (\ref{lar-thr-phr}) yields
\begin{equation}
\delta Q_{I}\! = \!
     \left( \!
            \begin{array}{cccc}
            \delta Q_{3}^{11}\!\! & \!\! 0 \!\! & \!\! 0 \!\! & \!\!
            \delta Q_{3}^{12}  \\
            0 \!\! & \!\! \delta Q_{4}^{22}\!\! & \!\! \delta
            Q^{21}_{4}\! \! & \!\! 0  \\
            0 \!\! & \!\! \delta Q^{12}_{4}\! \! & \!\! \delta
            Q_{4}^{11} \!\! & \!0  \\
            \delta Q^{21}_{3} \!\! & \!\! 0 \!\! & \!\! 0 \!\! & \!\! 
            \delta Q_{3}^{22}
            \end{array}
     \! \right) , \qquad
G_{0} = \frac{1}{2}
     \left( \!
            \begin{array}{cccc}
            G_{1}^{11}\!\! & \!\! G_{1}^{12} \!\!&\!\! 0\!\! &\!\! 0  \\
            G_{1}^{21}\!\! & \!\!G_{1}^{22}\!\! &\!\! 0\!\! &\!\! 0  \\
            0\!\! &\!\! 0 \!\!&\!\! G_{2}^{11} \!\!& \!\!G_{2}^{12}  \\
            0 \!\!& \!\!0 \!\!&\!\! G_{2}^{21}\!\! &\!\! G_{2}^{22}  \\
     \end{array}
    \! \right) ,
\label{deQI-G0-mat}
\end{equation}
where $\delta Q_{s}^{pp'}, G_{r}^{pp'}$ denote the matrices 
\begin{eqnarray} &&
\delta Q_{s}^{11} =
u_{s}^{1} \ ( \cos \theta_{s} - 1 )( u_{s}^{1} )^{-1}, \qquad
\delta Q_{s}^{12} =
i u_{s}^{1} \ \sin \theta_{s}
\mbox{e}^{i\phi_{s}} ( u_{s}^{2} )^{-1} ,
\nonumber \\ &&
\delta Q_{s}^{21} =
-i u_{s}^{2} \ \sin \theta_{s} 
\mbox{e}^{-i\phi_{s}} ( u_{s}^{1} )^{-1} , \qquad 
\delta Q_{s}^{22} =
u_{s}^{2} \ ( 1 - \cos \theta_{s} )( u_{s}^{2} )^{-1} , \qquad
\label{deQspp}
\end{eqnarray}
\begin{equation}
G_{r}^{12} = i \tan(\theta_{r} /2) \mbox{e}^{i\phi_{r}} ,
\qquad
G_{r}^{21} =  i \tan (\theta_{r} /2) \mbox{e}^{-i\phi_{r}} ,
\qquad
G_{r}^{pp} = 1 \ . \qquad
\label{Grpp}
\end{equation}

\subsection{Symmetries of ${\cal I}_{a}[ F ]$}
\label{sym-IaF}

The symmetry properties of our coset matrices imply symmetries of the
integrals $ {\cal I}_{a}[F] $. We use Eq.~(\ref{IaF-PKSD}) for ${\cal
  I}_{a}[F]$, the symmetry properties in Eqs.~(\ref{ULa-J12}) of
$U_{0}, \Lambda_{0}, U_{I}, \Lambda_{I}$, and we pass from the
coordinates $z$ to the coordinates $w$. We conclude that 
\begin{equation}
S_{j}( z_{12}, z_{34} ) = S_{j'}( w_{21}, w_{34} )
\label{Sj-J12}
\end{equation}
where $ j' = 1,2,4,3 $ for $ j = 1,2,3,4$ respectively.
With $ N_{j'} = N_{j} $ , this yields $ S( z_{12}, z_{34} ) = S(
w_{21}, w_{34} ) $. Similarly, we conclude that
\begin{equation}
D( z_{12}, z_{34} ) = D( w_{21}, w_{34} ) \ .
\label{D-J12}
\end{equation} 
>From Eq.~(\ref{rha}) we have $ \rho_{ a_{1} a_{2} a_{3} a_{4} }(
\theta_{12}, \theta_{34} ) = \rho_{ a_{2} a_{1} a_{3} a_{4} }(
\theta_{21}, \theta_{34} ) $, so that
\begin{eqnarray} &&
{\cal I}_{a_{1}a_{2}a_{3}a_{4}}[ F ] =
\int \mbox{d}[ w ]
\rho_{ a_{2} a_{1} a_{3} a_{4} }( \theta_{21}, \theta_{34} )
\nonumber \\ && \quad
\times \prod \nolimits^{(0)}_{q} \delta( w_{q} )
\prod \nolimits^{(1)}_{q} P( w_{q} ) \
K( \theta_{34} ) S( w_{21}, w_{34} ) D( w_{21}, w_{34} ) \ .
\label{IaF-J12}
\end{eqnarray}
The coordinates $ w_{2} $ are integrated over the same domain as the
coordinates $ w_{1} $, and the coordinates $ w_{q} $ over the same
domain as their counterparts $ z_{q} $. Therefore, the integral on the
r.h.s. of Eq.~(\ref{IaF-J12}) equals the integral $ {\cal I}_{ a_{2}
  a_{1} a_{3} a_{4}}[ F ] $. 
Using the symmetry properties in Eqs.~(\ref{ULa-J34}),
we can repeat the argument to show that
$ S(z_{12},z_{34}) = S(w_{12}, w_{43}) $ and $ D(z_{12},z_{34}) =
D(w_{12}, w_{43}) $. We recall that $ \rho_{ a_{1} a_{2} a_{3} a_{4}
  }( \theta_{12}, \theta_{34} ) = \rho_{ a_{1} a_{2} a_{4} a_{3} }(
\theta_{12}, \theta_{43} ) $ and find that in this case the coordinate
transformation yields the integral $ {\cal I}_{ a_{1} a_{2} a_{4}
  a_{3} }[ F ]. $ Combining these results shows that the integrals $
{\cal I}_{a} [ F ] $ satisfy the symmetry relations
\begin{equation}
{\cal I}_{ a_{1} a_{2} a_{3} a_{4} } [ F ] =
{\cal I}_{ a_{2} a_{1} a_{3} a_{4} } [ F ] = 
{\cal I}_{ a_{1} a_{2} a_{4} a_{3} } [ F ] =
{\cal I}_{ a_{2} a_{1} a_{4} a_{3} } [ F ] \ .
\label{IaF-J1234}
\end{equation}
These symmetry relations simplify the sum over the Efetov--Wegner
terms in Eq.~(\ref{IF-IaF}) to 
\begin{eqnarray} &&
{\cal I} [ F ] = {\cal I}_{1111} [ F ] + 2 {\cal I}_{1110} [ F ]
 + 2 {\cal I}_{1011} [ F ]   + {\cal I}_{1100} [ F ] 
\nonumber \\ && \qquad
+ 4 {\cal I}_{1010} [ F ] + {\cal I}_{0011}[ F ]
+ 2 {\cal I}_{1000} [ F ] + 2 {\cal I}_{0010} [ F ]
+ {\cal I}_{0000} [ F ]  \ . \qquad \qquad
\label{IF-IaF-sym}
\end{eqnarray}

\subsection{Vanishing $ {\cal I}_{a}[ F ] $ }
\label{van-IaF}

The sum in Eq.~(\ref{IF-IaF-sym}) simplifies even further: Due to the
vanishing of some of the $ {\cal I}_{a}[ F ] $'s, it reduces to four
terms. We first consider the case $ a = (1110) $. The integration
over $\delta( z_{4} )$ yields $ z_{4} = 0 $. Using
Eqs.~(\ref{deQI-G0-mat}) to (\ref{Grpp}) for $ \delta Q_{I} $ and $
G_{0} $, we find that 
\begin{equation}
V = \frac{1}{2} \left(
            \begin{array}{cccc}
            1   & x_{1}   & 0   &
            u^{1}_{3} \, x_{3} ( u^{2}_{3} )^{-1} y_{2} \\
            \bar x_{1}   & 1   & 0   & 0 \\
            0   & 0   & 1   & x_{2} \\
            u^{2}_{3} \, \bar x_{3} ( u^{1}_{3} )^{-1} y_{1} &
            0   & \bar x_{2}   & 1
            \end{array}
     \right),
\label{V/1110}
\end{equation}
and that
\begin{equation}
\mbox{detg}(1 + \delta W)
= (1-x_{3}^{f}\bar x_{3}^{f})(1-x_{3}^{b}\bar x_{3}^{b})^{-1} \ .
\label{detg/1110}
\end{equation}
Here,  $ x_{q} $, $ \bar x_{q} $ and $ y_{q} $ denote the matrices
\begin{equation} 
x_{q} = i \tan (\theta_{q} /2) \mbox{e}^{i\phi_{q}} \ ,
\bar x_{q} = i \tan (\theta_{q} /2) \mbox{e}^{-i\phi_{q}} \ ,
y_{q} = 1 - x_{q} \bar x_{q}  \ .
\label{xq-bxq-yq}
\end{equation}
The only part of $ V $ which contributes to
$I_{0}^{(rp)}VI_{0}^{(r'p')}$ is the block $V_{rp,r'p'}$. According to
Eq.~(\ref{V/1110}), $ V_{12,21} = V_{21,12} = 0 $, and the only
nonzero part of $ S $ is the term $ N_{2} S_{2} $. However, $ S_{2} $
contains neither $\gamma_{3}^{p}$ nor $\gamma_{3}^{p\ast}$, and 
Eq.~(\ref{detg/1110}) shows that the same is true for the function $ D
$. Thus $P(z_{3})SD = 0 $, and the integral ${\cal I}_{1110}[F] $ 
is equal to zero. For $ a = (1010) $, $z_{2}=z_{4}=0$ and, therefore,
$ V_{12,21} = V_{21,12} = V_{21,22} = V_{22,21} = 0 $. Thus $S = 0$,
and $ {\cal I}_{1010}[ F ] = 0 $. For the same reason, $ {\cal
  I}_{1000}[ F ] = {\cal I}_{0010}[ F ] = {\cal I}_{0000}[ F ] = 0
$. The sum in Eq.~(\ref{IF-IaF-sym}) simplifies to 
\begin{equation}
{\cal I}[ F ]
= {\cal I}_{1111}[ F ] + 2 {\cal I}_{1011}[ F ]
+ {\cal I}_{1100}[ F ] + {\cal I}_{0011}[ F ]  \ .
\label{IF-IaF-fin}
\end{equation}
The term $ {\cal I}_{1100}[ F ] $ is independent of $ t $ and yields
the limit $ t \rightarrow \infty $ of $ {\cal I}[ F ] $. For this
term, $K=1$, $D=D_{0}$, and
$ S=N_{2}S_{2} =
N_{2}\prod_{rp} \gamma_{r}^{p}\gamma_{r}^{p\ast} \ \prod_{r}<
x_{r}\bar x_{r} > $. Integrating over $\phi_{r}, \gamma_{r}$ and
$\gamma_{r}^{\ast}$ we are left with 
\begin{equation}
I_{1100}[ F ] = ( M^{4} / 4 )
\int \prod_{r} \mbox{d} [ \theta_{r} ] \
\prod_{r} \rho_{ \mbox{\scriptsize G} } ( \theta_{r} )
< \tan^{2}(\theta_{r}/2) > \ D_{0} = M^{2}/4 \ .
\label{IaF/1100}
\end{equation}
This is the disconnected part of $\overline{g^{(1)}g^{(2)}}$. Thus,
the autocorrelation function $C(t,M)$ is given by
\begin{equation}
C(t,M)
= {\cal I}_{1111}[ F ] + 2 {\cal I}_{1011}[ F ]
+ {\cal I}_{0011}[ F ]  \;.
\label{deg12-IaF}
\end{equation}

\section{ The Limit of Large $ t $ at Fixed $ M $ }
\label{t>>1}

In this section, we expand the autocorrelation function $C(t,M)$
for fixed $M$ asymptotically in inverse powers of $t$, and evaluate
the leading term. In the limit $t \gg 1$, the integrals ${\cal I}_{a}[
F ]$ appearing on the r.h.s. of Eq.~(\ref{deg12-IaF}) are dominated by
the contribution of the neighbourhood of the surface $
\theta_{s}^{\alpha} = 0 $ where $ Q_{I} = L $, and where the graded
trace in the exponent of the coupling term
\begin{equation}
K = \mbox{e}^{ -( t/2 ) < ( Q_{I} \tau_{3} )^{2} > }
\label{K-QI}
\end{equation}
is equal to zero.

\subsection{ Asymptotic Expansion}
\label{asy-exp/t}

We start from Eq.~(\ref{IaF-PKSD}), introduce the rescaled angles
$ \tilde \theta_{s}^{\alpha} = t^{1/2} \theta_{s}^{\alpha} $, and pass
from the coordinates $ z_{s} $ to the coordinates $\tilde z_{s} = (
\tilde \theta_{s}^{b}, \phi_{s}^{b}, \tilde \theta_{s}^{f},
\phi_{s}^{f}, \gamma_{s}^{1}, \gamma_{s}^{1\ast}, \gamma_{s}^{2},
\gamma_{s}^{2\ast} )$ with $ \mbox{d}[ z_{s} ] = \mbox{d} [ \tilde
z_{s} ] t^{-1} $ and $ P( z_{s} ) = P( \tilde z_{s} ) $. We expand
$ t^{-2} \rho_{a} $, $ K $, $ V $ and $ D $ in powers of $t^{-1/2}$,
\begin{eqnarray} &&
t^{-2} \rho_{a}
= \sum_{n_{\rho}} t^{-n_{\rho}/2}
\rho_{a}^{(n_{\rho})} \ ,
\qquad
K = \sum_{n_{k}} t^{-n_{k}/2}
K^{ ( n_{k} ) } \ ,
\nonumber \\ &&
V I_{0}^{(rp)}
= \sum_{n_{rp}} t^{-n_{rp}/2}
V^{(n_{rp})} I_{0}^{(rp)} \ ,
\qquad
D = \sum_{n_{d}} t^{-n_{d}/2}
D^{ ( n_{d} ) } \ .
\label{rhaKVD-n/t}
\end{eqnarray}
The coefficients $\rho_{a}^{(n_{\rho})}, K^{(n_{k})}, V^{(n_{rp})},
D^{(n_{d})}$ are functions of $ z_{r},\tilde z_{s} $. Collecting terms
of the same order in $ t^{-1/2} $, we get
\begin{equation}
{\cal I}_{a}[ F ] = \sum_{n} t^{-n/2} {\cal I}_{a}^{(n)}[ F ] \ .
\label{IaF-IanF/t}
\end{equation}
Here ${\cal I}_{a}^{(n)}[ F ]$ denotes the sum of integrals
\begin{equation}
{\cal I}_{a}^{(n)}[ F ]
= \sum_{ j} \sum_{ n_{\rho} n_{k} n_{11} n_{12} n_{21} n_{22} n_{d} }
{\cal I}_{a}^{ ( n_{\rho}; n_{k}; n_{v}; n_{d} ) }[ F_{j} ] \
\label{IanF-IanFj/t}
\end{equation}
with
\begin{eqnarray} &&
{\cal I}_{a}^{ ( n_{\rho}; n_{k}; n_{v}; n_{d} ) }[ F_{j} ]
= \int \prod \nolimits_{r} \mbox{d}[ z_{r} ] 
\prod \nolimits_{s} \mbox{d}[ \tilde z_{s} ]
\rho_{a}^{ ( n_{\rho} ) }
\nonumber \\ && \qquad \times
\prod \nolimits_{r}^{(0)} \delta ( z_{r} )
\prod \nolimits_{r}^{(1)} P( z_{r} ) \prod \nolimits_{s} P( \tilde z_{s} )
K^{ ( n_{k} ) } N_{j} S_{j}^{(n_{v})} D^{ ( n_{d} ) } \ . \qquad
\label{IanFj/t}
\end{eqnarray}
Here, $ n_{v} = ( n_{11} n_{12} n_{21} n_{22} ) $, 
and $ S_{j}^{(n_{v})} $ are given by Eqs.~(\ref{Sj-VI0}) with 
$ V I_{0}^{(rp)} $ replaced by $ V^{(n_{rp})} I_{0}^{(rp)} $ ,
\begin{eqnarray} &&
S_{1}^{ ( n_{v} ) } =
\, < V^{ (n_{11}) } I_{0}^{(11)} V^{ (n_{22}) } I_{0}^{(22)} > \
< V^{ (n_{12}) } I_{0}^{(12)} V^{ (n_{21}) } I_{0}^{(21)} >,
\nonumber \\ &&
S_{2}^{ ( n_{v} ) } =
\, < V^{ (n_{11}) } I_{0}^{(11)} V^{ (n_{12}) } I_{0}^{(12)} > \
< V^{ (n_{21}) } I_{0}^{(21)} V^{ (n_{22}) } I_{0}^{(22)} > \ ,
\nonumber \\ &&
S_{3}^{ ( n_{v} ) } =
\, < V^{ (n_{11}) } I_{0}^{(11)} V^{ (n_{22}) } I_{0}^{(22)}
V^{ (n_{21}) } I_{0}^{(21)} V^{ (n_{12}) } I_{0}^{(12)} > \ ,
\nonumber \\ &&
S_{4}^{ ( n_{v} ) } =
\, < V^{ (n_{11}) } I_{0}^{(11)} V^{ (n_{12}) } I_{0}^{(12)}
V^{ (n_{21}) } I_{0}^{(21)} V^{ (n_{22}) } I_{0}^{(22)} > \ .
\label{Sjnv-VnI0}
\end{eqnarray}
The order of $ S_{j}^{(n_{v})} $ is $n_{\sigma} = \sum_{rp}n_{rp}$.  
The sum over $ n_{\rho}, n_{k}, n_{rp} $ and $ n_{d} $ is restricted
by the condition $ n_{\rho} + n_{k} + n_{\sigma} + n_{d} = n $. Since
$ K $ is given by the exponential (\ref{K-QI}), all integrals contain
the exponential factor
\begin{equation}
K^{(0)} =
\mbox{e}^{ 2 \sum_{s} < ( \tilde \theta_{s} )^{2} > } \ .
\label{K0/t}
\end{equation}
On extending the domain of integration region over $\tilde
\theta_{s}^{f}$ from zero to infinity, the series (\ref{IaF-IanF/t})
yields an asymptotic expansion for ${\cal I}_{a}[ F ]$ . Since
$\rho_{a}$, $K$, $S$ and $D$ are even functions of
$\theta_{s}^{\alpha}$, only terms with even $n$ appear, and the 
expansion proceeds in inverse powers of $t$. Thus,
\begin{equation}
C(t,M)
= \sum_{n=1}^{\infty} t^{-n} \bigg(\ {\cal I}_{1111}^{(2n)}[ F ]
   + 2 {\cal I}_{1011}^{(2n)}[ F ]
   + {\cal I}_{0011}^{(2n)}[ F ] \  \bigg) \;.
\label{deg12-asyexp/t}
\end{equation}
To calculate the expansion coefficients $ {\cal I}_{a}^{ ( 2n ) } [ F(
Q ) ] $ , we use the symmetry properties of $ U_{0}, \Lambda_{0},
U_{I}, \Lambda_{I} $. It follows (see Appendix~\ref{sym-pro}) that the
integrals $ {\cal I}_{a}^{ ( n_{\rho}; n_{k}; n_{v}; n_{d} ) }[ F_{j}
] $ satisfy the symmetry relations
\begin{eqnarray} &&
{\cal I}_{ a_{1} a_{2} 11 }
^{ ( n_{\rho}; n_{k}; n_{11} n_{12} n_{21} n_{22}; n_{d} ) } [ F_{j} ]
= {\cal I}_{ a_{2} a_{1} 11 }
^{ ( n_{\rho}; n_{k}; n_{22} n_{21} n_{12} n_{11}; n_{d} ) } [ F_{j'} ]
\nonumber \\ && 
\qquad = {\cal I}_{ a_{1} a_{2} 11 }
^{ ( n_{\rho}; n_{k}; n_{12} n_{11} n_{22} n_{21}; n_{d} ) } [ F_{j'} ] 
= {\cal I}_{ a_{2} a_{1} 11 }
 ^{ ( n_{\rho}; n_{k}; n_{21} n_{22} n_{11} n_{12}; n_{d} ) } 
[ F_{j} ] \ , \qquad
\label{IanFj-sym/t}
\end{eqnarray}
where $ j' = 1,2,4,3 $ for $ j = 1,2,3,4 $, respectively.

To find the leading term of the expansion we recall two properties of 
${\cal I}_{a}^{(n_{\rho};n_{k};n_{v};n_{d})}[F_{j}]$: (i) This expression
is nonzero only when $ \sum_{s} P(z_{s}) S_{j}^{(n_{v})} D^{(n_{d})} $
is nonzero and (ii) the expression does not vanish when integrated
over the phases $ \phi_{q}^{\alpha} $. Inspecting the explicit
expressions of the relevant low--order terms $ S_{j}^{(n_{v})}
D^{(n_{d})} $ as given by the matrices $ ( \delta Q_{s}^{pp'})^{(n)} $
and $ G_{r}^{pp'}$, we find that these conditions are first met for
$n_{\sigma} + n_{d} = 4 $. Thus, the leading term $C_{ t \gg 1 }(t,M)$
of the expansion is of order $ t^{-2} $ and given by 
\begin{equation}
C_{ t \gg 1 }(t,M)
   = t^{-2} \bigg( \ {\cal I}_{1111}^{(4)}[ F ]
   + 2 {\cal I}_{1011}^{(4)}[ F ]
   + {\cal I}_{0011}^{(4)}[ F ] \ \bigg) \ .
\label{deg12-leater/t}
\end{equation}
For each of the ${\cal I}_{a}^{(4)}[F]$ in Eq.~(\ref{deg12-leater/t}),
the sum on the r.h.s. of Eq.~(\ref{IanF-IanFj/t}) reduces to the terms
with $n_{\rho}=n_{k}=0$.

\subsection{ The Leading Term $C_{t \gg 1}(t,M)$ }
\label{lea-ter/t}

We first calculate the volume integral 
\begin{equation}
{\cal I}_{1111}^{(4)}[ F ] =
 \sum_{j} \sum_{ n_{11} n_{12} n_{21} n_{22} n_{d} }
{\cal I}_{1111}^{(0;0;n_{v};n_{d})}[ F_{j} ] \ .
\label{IanF-IanFj/volt}
\end{equation}
We suppress the tildes above $ z_{s} $ and $\theta_{s} $. Inspection
of the explicit expressions of the relevant terms $ S_{j}^{(n_{v})}
D^{(n_{d})} $ shows that the nonvanishing contributions to $
\sum_{s}P(z_{s}) S_{j}^{(n_{v})} D^{(n_{d})} $ stem solely from the
terms which are linear in all four matrices $( \delta Q_{s}^{12}
)^{(1)}$, $ ( \delta Q_{s}^{21} )^{(1)} $ and which for each $ r $  
contain the same number of $ G_{r}^{12} $ and of $ G_{r}^{21} $. We
restrict the sum over ${\cal I}_{1111}^{(0;0;n_{v};n_{d})}[F_{j}]$ 
accordingly, employ the symmetry relations  Eq.~(\ref{IanFj-sym/t}),
and find
\begin{eqnarray} &&
{ \cal I }_{1111}^{(4)} [ F ]
=   { \cal I }_{1111}^{(0;0;1111;0)} [ F_{1} ]
\nonumber \\ && \quad
+   { \cal I }_{1111}^{(0;0;0000;4)} [ F_{2} ]
+ 4 { \cal I }_{1111}^{(0;0;0002;2)} [ F_{2} ]
+ 2 { \cal I }_{1111}^{(0;0;0022;0)} [ F_{2} ]
\nonumber \\ && \quad
+ 2 { \cal I }_{1111}^{(0;0;0220;0)} [ F_{2} ]
+ 2 { \cal I }_{1111}^{(0;0;0101;2)} [ F_{3} ]
+ 4 { \cal I }_{1111}^{(0;0;0121;0)} [ F_{3} ] \ . \qquad
\label{IanF-IanFj-sym/volt}
\end{eqnarray}
The integrals ${\cal I}_{1111}^{(0;0;n_{v};n_{d})}[F_{j}]$ have the
form shown in Eq.~(\ref{IanFj/t}). Explicit expressions in terms of
polar coordinates for the $ V^{(n)}, D^{(n)} $ contributing to
Eq.~(\ref{IanF-IanFj-sym/volt}) are given in Apendix~\ref{IanF-app/t}.
Substituting these formulae and applying the projectors $P(z_{r})$
yields
\begin{eqnarray} &&
{\cal I}_{1111}^{(4)}[ F ] =
(M^{2}/64) \int \mbox{d}[ z ]
\rho_{1111}^{(0)}
K^{(0)} D_{0} \prod \nolimits_{rp} \gamma^{p}_{r} \gamma^{p\ast}_{r} P
\nonumber \\ && \quad \quad
\times \big(\
2 < X_{3}^{(1)} y_{2} \bar X_{3}^{(1)} y_{1} >
  < X_{4}^{(1)} y_{1} \bar X_{4}^{(1)} y_{2} >
\nonumber \\ && \quad \quad \quad
+ M^{4} < y_{1} >< y_{2} >
< X_{3}^{(1)} \bar x_{2} X_{4}^{(1)} \bar x_{1} >
< \bar X_{3}^{(1)} x_{1} \bar X_{4}^{(1)} x_{2} >
\nonumber \\ && \quad \quad \quad \quad
+ 4 M^{3} < y_{1} >
< X_{3}^{(1)} \bar x_{2} y_{2} X_{4}^{(1)} \bar x_{1} >
< \bar X_{3}^{(1)} x_{1} \bar X_{4}^{(1)} x_{2} >
\nonumber \\ && \quad \quad \quad \quad \quad
- 2 M^{2} < y_{1} >
< X_{3}^{(1)} y_{2} \bar X_{3}^{(1)} x_{1}
\bar X_{4}^{(1)} y_{2} X_{4}^{(1)} \bar x_{1} >
\nonumber \\ && \quad \quad \quad \quad \quad \quad
+ 2 M^{2}
< X_{3}^{(1)} \bar x_{2} X_{4}^{(1)} \bar x_{1} y_{1} >
< \bar X_{3}^{(1)} x_{1} \bar X_{4}^{(1)} x_{2} y_{2} >
\nonumber \\ && \quad \quad \quad \quad \quad \quad \quad
+ 2 M^{2}
< X_{3}^{(1)} \bar x_{2} y_{2} X_{4}^{(1)} \bar x_{1} y_{1} >
< \bar X_{3}^{(1)} x_{1} \bar X_{4}^{(1)} x_{2} >
\nonumber \\ && \quad \quad \quad \quad \quad \quad \quad \quad
- 4 M 
< X_{3}^{(1)} y_{2} \bar X_{3}^{(1)} x_{1}
\bar X_{4}^{(1)} y_{2} X_{4}^{(1)} \bar x_{1} y_{1} >
\quad \quad \big) \ , \qquad \qquad
\label{IanF-Xnxn/volt}
\end{eqnarray}
where $X_{s}^{(1)}, \bar X_{s}^{(1)}$ denote the matrices
\begin{eqnarray} &&
X_{s}^{(1)} = ( \delta Q_{s}^{12} )^{(1)}
= i u_{s}^{1} \ \theta_{s}\mbox{e}^{i\phi_{s}} (u_{s}^{2})^{-1} \ ,
\nonumber \\ &&
\bar X_{s}^{(1)} = ( \delta Q_{s}^{21} )^{(1)}
= - i u_{s}^{2} \ \theta_{s} \mbox{e}^{-i\phi_{s}} (u_{s}^{1})^{-1} \ ,
\label{Xs1-bXs1}
\end{eqnarray}
and where 
\begin{equation}
x_{r} = G_{r}^{12}\ , \qquad  \bar x_{r} = G_{r}^{21} \ ,
\qquad
 y_{r} = 1 - x_{r} \bar x_{r} 
\label{xr-bxr-yr}
\end{equation}
are the matrices introduced in Eq.~(\ref{xq-bxq-yq}). The density
$\rho_{1111}^{(0)}$ is given by
\begin{equation}
\rho_{1111}^{(0)} = 
\prod_{r} \rho_{ \mbox{\scriptsize G} }( \theta_{r} )
\prod_{s}\rho_{ \mbox{\scriptsize G} }^{(0)}( \theta_{s} ) \ ,
\qquad
\rho_{ \mbox{\scriptsize G} }^{(0)}( \theta_{s} ) =
4 \prod_{\alpha} \theta_{s}^{\alpha}\, < ( \theta_{s} )^{2} >^{-2} \ ,
\label{rh0/volt}
\end{equation}
and $P$ denotes the projector $P = \prod_{s}P(z_{s})$ which we
suppress in the sequel. First we integrate over $ \phi_{s},
\gamma_{s}, \gamma_{s}^{\ast}$. This can be done making use of
\begin{eqnarray} &&
\int \prod_{s} \mbox{d}[ \chi_{s} ] 
< X_{3}^{(1)} \, A_{I} \, \bar X_{3}^{(1)} \, B_{I} >
< X_{4}^{(1)} \, A_{II} \, \bar X_{4}^{(1)} \, B_{II} > \qquad
\nonumber \\ &&
\qquad\qquad = \prod_{s} < ( \theta_{s} )^{2} >
 \ < A_{I} >< B_{I} >< A_{II} >< B_{II} >  \;,
\nonumber \\ &&
\int \prod_{s} \mbox{d} [ \chi_{s} ]
< X_{3}^{(1)} \, A_{I}\, X_{4}^{(1)} \, B_{I} >
< \bar X_{3}^{(1)} \, A_{II}\, \bar X_{4}^{(1)} \, B_{II} >\qquad
\nonumber \\ &&
\qquad\qquad = \prod_{s} < ( \theta_{s} )^{2} >
\ < A_{I} B_{II} >< B_{I} A_{II}>  \;,
\nonumber \\ &&
\int \prod_{s} \mbox{d} [ \chi_{s} ]
< X_{3}^{(1)} \, A_{I}\, \bar X_{3}^{(1)} \, B_{I} \,
\bar X_{4}^{(1)} \, A_{II}\, X_{4}^{(1)} \, B_{II} >\qquad
\nonumber \\ &&
\qquad\qquad  = \prod_{s} < ( \theta_{s} )^{2} >
\ < A_{I} >< A_{II} >< B_{I} B_{II}> \ .
\label{int-ch}
\end{eqnarray}
These formulae are valid for all two--dimensional graded matrices
$A_{I},B_{I},$ ~$A_{II},B_{II}$ which are independent of $ \phi_{s},
\gamma_{s}, \gamma_{s}^{\ast}$, and can be verified with the help of
\begin{equation}
\int \mbox{d} \gamma_{s}^{p} \mbox{d} \gamma_{s}^{p\ast}
( u_{s}^{p} )_{ \alpha_{1} \alpha_{2} }
( u_{s}^{p} )_{ \alpha_{1}^{\prime} \alpha_{2}^{\prime} }^{-1}
= (-)^{ p + \alpha_{2} }
\delta_{ \alpha_{1} \alpha_{2}^{\prime} }
\delta_{ \alpha_{2} \alpha_{1}^{\prime} }
( 2 \pi )^{-1} \ .
\label{int-gas}
\end{equation}
Integrating over $\phi_{r},\gamma_{r},\gamma_{r}^{\ast}$ then
simplifies the volume term to an ``eigenvalue'' integral over the
angles $ \theta_{q}^{\alpha} $ ,
\begin{eqnarray} &&
{\cal I}_{1111}^{(4)}[ F ] = (M^{2}/64)
   \int \prod \nolimits_{q} \mbox{d}[\theta_{q}]
   \rho_{1111}^{(0)}
    K^{(0)} D_{0} \prod \nolimits_{s} < (\theta_{s})^{2} >
\nonumber \\ && \qquad
\times \bigg(\! < y_{1} >^{2}\Big( (M^{4}+2M^{2}+2)\!< y_{2} >^{2}\!
- 4M^{3}< y_{2} ( 1 - y_{2} ) \!> \Big)
\nonumber \\ && \qquad \qquad
- 4M\! < y_{1} ( 1 - y_{1} ) >\! \Big( < y_{2} >^{2}\!
- M \! < y_{2} ( 1 - y_{2} ) >\! \Big) \bigg) \ . \qquad\quad
\label{IanF-th/volt}
\end{eqnarray}
Since the exponentials $ K^{(0)}, D_{0} $ factorize into products of
exponentials each depending on one of the $\theta_{q}$, the
integration over angles simplifies to the evaluation of
two--dimensional integrals over $ \theta_{q} $. The integrals over $
\theta_{r} $ can be done as in Ref.~\cite{gos98}. With
\begin{eqnarray} &&
\int \mbox{d} [ \theta_{r} ]
\rho_{ \mbox{\scriptsize G} } ( \theta_{r} )
\mbox{e}^{ -M <\ln (1 + \cos\theta_{r})> }
< y_{r} >^{2}\, = ( 1 - M^{2} )^{-1}  \;,
\nonumber \\ &&
\int \mbox{d} [ \theta_{r} ]
\rho_{ \mbox{\scriptsize G} } ( \theta_{r} )
\mbox{e}^{ -M <\ln (1 + \cos\theta_{r})> }
< y_{r} ( 1 - y_{r} ) > \,
= M ( 1 -  M^{2} )^{-1} 
\label{int-thr/t}
\end{eqnarray}
and
\begin{eqnarray} &&
\int \mbox{d} [ \theta_{s} ]
\rho_{ \mbox{\scriptsize G} }^{(0)} ( \theta_{s} )
\mbox{e}^{2<(\theta_{s})^{2}>} < (\theta_{s})^{2} > 
\nonumber \\ && \quad
= 4 \int_{0}^{i\infty} \mbox{d} \theta_{s}^{b}
\int_{0}^{\infty} \mbox{d} \theta_{s}^{f}
\mbox{e}^{ 2 < ( \theta_{s} )^{2} > } 
\theta_{s}^{b} \theta_{s}^{f}
< ( \theta_{s} )^{2} >^{-1}
\ = 1/2 \ ,
\label{int-ths/t}
\end{eqnarray}
the result is
\begin{equation}
\ {\cal I}_{1111}^{(4)}[ F ]
= \frac{M^{2}}{256}\, \frac{ M^{4} -2 M^{2} + 2 }{( M^{2} - 1 )^{2}} \ .
\label{IanF-res/volt}
\end{equation}
The boundary terms can be evaluated in the same way, see
Appendix~\ref{IanF-app/t}. For $ a = (1011) $ we have $z_{2}=0$, which
reduces the number of contributing integrals. We find
\begin{eqnarray}
{ \cal I }_{1011}^{(4)} [ F ]
&=& { \cal I }_{1011}^{(0;0;1111;0)} [ F_{1} ]
+   { \cal I }_{1011}^{(0;0;0022;0)} [ F_{2} ]
+ 2 { \cal I }_{1011}^{(0;0;0121;0)} [ F_{3} ]
\nonumber \\
&=& -\frac{M^{2}}{256}\frac{ M^{2} - 2 }{ M^{2} - 1 } \ .
\label{IanF-res/bou1t}
\end{eqnarray}
For $ a = (0011) $ we have $ z_{1}=z_{2}=0 $, and only one integral
contributes,
\begin{equation} 
{ \cal I }_{0011}^{(4)} [ F ] =
{ \cal I }_{0011}^{(0;0;1111;0)} [ F_{1} ] 
= \frac{M^{2}}{128} \ .
\label{IanF-res/bou2t}
\end{equation}
We thus find that the leading term is
\begin{eqnarray}
C_{ t \gg 1 }(t,M)
&=& t^{-2} \Big( \ {\cal I}_{1111}^{(4)}[ F ]
+ 2 {\cal I}_{1011}^{(4)}[ F ]
+ {\cal I}_{0011}^{(4)}[ F ] \ \Big) 
\nonumber \\ 
&=& \frac{1}{4}\frac{ M^{4} }{(M^{2} - 1)^{2} }
\left( \frac {M}{8t} \right)^{2} \ . \qquad 
\label{lea-ter-res/t}
\end{eqnarray}
This expression is discussed in the next Section \ref{conclusions}.

\section{Summary and Conclusions}
\label{conclusions}

We have investigated the magnetoconductance autocorrelation function
for ballistic electron transport through microstructures having the
form of a classically chaotic billiard. The structures were assumed to
be connected to ideal leads carrying few channels. Assuming ideal
coupling between leads and billiard, we have described the system in
terms of a random matrix model.

The autocorrelation function depends only on the field parameter $t$,
specified by the field (flux) difference, and on the channel number
$M$. Using the supersymmetry technique, we have calculated analytically
the leading terms in the asymptotic expansion for large $t$ at fixed
$M$, and, in Appendix \ref{lea-ter/M}, for large $M$ at fixed $t/M$. 
To this end, we have employed a new parametrization of the coset space
which is particularly convenient for the present problem of partly
broken symmetry. Applying Berezin's theory of coordinate
transformations in superintegrals, we succeeded in formulating the
relevant integral theorem, and in identifying volume and boundary
(Efetov--Wegner) terms. Symmetry properties of the coset matrices
helped us in considerably simplifying the evaluation of these terms. 
We believe that the method developed in this paper is of general
interest for the supersymmetry technique, and we hope it will be
helpful in other cases. We have shown that the Efetov--Wegner terms
are essential for details of the large $t$ behaviour at small $M$.

To discuss our results, we construct a function $C_{SL}^{I}(t,M)$ of
the form of a squared Lorentzian which has the same leading term at
large $t$ as our result~(\ref{lea-ter-res/t}). We use the value of the
autocorrelation function at the origin $t = 0$ given by the variance
$M^{2}/[4(M^{2}-1)]$ of the dimensionless conductance $g(B)$, see
Ref.~\cite{bar94}. This yields
\begin{equation}
C_{SL}^{I}(t,M) 
= \frac{M^{2}}{ 4(M^{2} - 1) } \ 
\Bigg( \ 1 + \frac{8t}{M^{2}} \ \sqrt{{M^{2}} - 1 } \ \Bigg)^{-2} \ .
\label{CSLI}
\end{equation}
In our earlier paper \cite{gos98}, we found that the leading terms in
the asymptotic expansion of the autocorrelation function for small $t$
are
\begin{eqnarray} &&
C(t,M) = \frac{M^{2}}{4(M^{2}-1)}
\, - \, \frac{4M^{3}}{(M^{2}-1)^{2}} \, t
\nonumber \\ && \qquad
+ \, \frac{ 16M^{2}\, (3M^{4}-11M^{2}+36)}
{(M^{2}-1)^{2} \, (M^{2}-4)\, (M^{2}-9)} \, t^{2}\, + \, O(t^{3}) \ .
\label{Csmt}
\end{eqnarray}
The squared Lorentzian consistent with the first two terms is
\begin{equation}
C_{SL}^{II}(t,M) 
= \frac{M^{2}}{ 4(M^{2} - 1) } \ 
\Bigg(  \ 1 + \frac{8t}{M} \ \frac{M^{2}}{M^{2}-1} \ \Bigg)^{-2} \ .
\label{CSLII}
\end{equation}
The two squared Lorentzians are different. This shows that we cannot
find a function with the shape of a squared Lorentzian which has the
same leading terms as the exact correlation function for both small
and large values of $t$. In other words, the exact shape of the
autocorrelation function is not a squared Lorentzian. However, with
increasing $M$, the differences between $C_{SL}^{I}(t,M)$ and
$C_{SL}^{II}(t,M)$ decrease, and for $M \gg 1$, the semiclassical
result  
\begin{equation} 
C_{SL}(t,M) = \frac{1}{4}\Bigg( \ 1 + \frac{8t}{M} \ \Bigg)^{-2}
\label{CSL}
\end{equation}
does have the leading terms required by Eqs.~(\ref{lea-ter-res/t}) and
(\ref{Csmt}). Moreover, the numerical differences between the three
functions $C_{SL}, C_{SL}^{I}$ and $C_{SL}^{II}$ are small even at
small $M$.
>From the numerical simulations discussed in Ref.~\cite{gos98} 
the difference between the exact RMT autocorrelation function
and the best squared Lorentzian fit does not exceed $5\%$ in
magnitude even for the smallest value of $M$, $M=2$. 

\subsection*{Acknowledgments}

Z.P. thanks the members of the Max-Planck-Institut f\"ur Kernphysik in
Heidelberg for their hospitality and support, and acknowledges support
by the Grant Agency of Czech Republic (grant 202/96/1744).

\section{Appendices}
\label{appendices}

\subsection{Symmetry Properties of the Integrals \\ 
$ {\cal I}_{a_{1} a_{2} 1 1}
^{ ( n_{\rho}; n_{k}; n_{v}; n_{d} ) } [ F_{j} ] $}
\label{sym-pro}

Similar to the symmetry properties of $ I_{a}[F] $ discussed in
Subsection \ref{sym-IaF}, these properties follow from the symmetries
of the matrices $U_{0},\Lambda_{0},U_{I},\Lambda_{I}$ under the
similarity transformations by the permutation matrices $J_{12}$ and
$J_{34}$, and can be found in the same way. The integrals are given by
Eq.~(\ref{IanFj/t}). We use Eqs.~(\ref{ULa-J12}) and change from the
coordinates $ z_{r}, \tilde z_{s} $ to the coordinates $ w_{r}, \tilde
w_{s} $, where $ \tilde w_{s} = ( \tilde \theta_{r}^{b}, \psi_{r}^{b},
\tilde \theta_{r}^{f}, \psi_{r}^{f}, \beta_{s}^{1}, \bar \beta_{s}^{1}, 
\beta_{s}^{2}, \bar \beta_{s}^{2} )$. Since the expansion terms
satisfy the same relations as the expanded functions, we have 
\begin{equation}
S_{j}^{(n_{11} n_{12} n_{21} n_{22})}( z_{12}, \tilde z_{34} ) =
S_{j'}^{(n_{22} n_{21} n_{12} n_{11})}( w_{21}, \tilde w_{34} )
\label{Sjn-J12/t}
\end{equation}
with $j' = 1,2,4,3$ for $j=1,2,3,4$, respectively. Similarly,
\begin{equation}
D^{(n_{d})}(z_{12},\tilde z_{34}) =
D^{(n_{d})}(w_{21},\tilde w_{34}) \ .
\label{Dn-J12}
\end{equation}
With $ \rho_{ a_{1} a_{2} 11}^{(n_{\rho})}( \theta_{12}, \tilde
\theta_{34} ) = \rho_{ a_{2} a_{1} 11}^{(n_{\rho})}( \theta_{21},
\tilde \theta_{34} ) $ we finally find that
\begin{eqnarray} &&
{\cal I}_{ a_{1} a_{2} 11 }
^{ ( n_{\rho}; n_{k}; n_{11} n_{12} n_{21} n_{22}; n_{d} ) }
[ F_{j} ]
\nonumber \\ && \quad
= \int \prod \nolimits_{r} \mbox{d}[ w_{r} ] \ 
\prod \nolimits_{s} \mbox{d}[ \tilde w_{s} ] \
\rho_{ a_{2} a_{1} 11}^{ (n_{\rho}) }
( \theta_{21}, \tilde \theta_{34} )
\prod \nolimits_{r}^{(0)} \delta ( w_{r} )
\prod \nolimits_{r}^{(1)} P( w_{r} )
\nonumber \\ && \quad
\times 
\prod \nolimits_{s} P( \tilde w_{s} )
K^{ ( n_{k} ) }( \tilde \theta_{34} )
N_{j'} S_{j'}^{ ( n_{22} n_{21} n_{12} n_{11} ) }( w_{21}, \tilde w_{34} )
D^{ ( n_{d} ) }( w_{21}, \tilde w_{34} ). \qquad \quad
\label{IanFj-J12/t}
\end{eqnarray}
Since the coordinates $ w_{2} $ are integrated over the same domain as
$ w_{1} $, and the coordinates $ w_{r}, \tilde w_{s} $ over the same
domain as $ z_{r}, \tilde z_{s} $, the integral on the r.h.s. of
Eq.~(\ref{IanFj-J12/t}) is the integral $ {\cal I}_{ a_{2} a_{1} 11 }
^{ ( n_{\rho}; n_{k}; n_{22} n_{21} n_{12} n_{11}; n_{d} ) } [ F_{j'}
] $ . The same procedure can be repeated using the symmetry properties
with respect to $ J_{34} $ instead of those with respect to $J_{12}$.
In this case, the coordinate transformation yields the integral 
$ {\cal I}_{ a_{1} a_{2} 11 }^{ ( n_{\rho}; n_{k}; n_{12} n_{11}
  n_{22} n_{21} ; n_{d} ) }[ F_{j'} ] $. Summarizing and combining the
results we find that the integrals $ {\cal I}_{ a_{1} a_{2} 11 }
^{ ( n_{\rho}; n_{k}; n_{r}; n_{d} ) }[ F_{j} ] $ satisfy the symmetry
relations in Eqs.~(\ref{IanFj-sym/t}).

\subsection{The Integrals $ {\cal I}_{a}^{(4)}[ F ] $ }
\label{IanF-app/t}

For the volume term ${\cal I}_{1111}^{(4)}[F]$, the functions
$V^{(1)},V^{(2)},D^{(2)},D^{(4)}$ take the form 
\begin{displaymath}
V^{(1)} = \frac{1}{4}
     \left(
          \begin{array}{cccc}
          0 & 0 & 0 & X_{3}^{(1)} y_{2} \\
          0 & 0 & - \bar X_{4}^{(1)} y_{2} & 0 \\
          0 & X_{4}^{(1)} y_{1} & 0 & 0 \\
          - \bar X_{3}^{(1)} y_{1} & 0 & 0 & 0
          \end{array}
     \right) \ ,
\end{displaymath}
\begin{displaymath}
V^{(2)} = - \frac{1}{8}
     \left(
          \begin{array}{cccc}
          0 & X_{3}^{(1)} \bar x_{2} X_{4}^{(1)} y_{1} & 0 & 0 \\
          \bar X_{4}^{(1)} x_{2} \bar X_{3}^{(1)} y_{1} & 0 & 0 & 0 \\
          0 & 0 & 0 & X_{4}^{(1)} \bar x_{1} X_{3}^{(1)} y_{2} \\
          0 & 0 & \bar X_{3}^{(1)} x_{1} \bar X_{4}^{(1)} y_{2} & 0
          \end{array}
     \right) \ ,
\end{displaymath}
\begin{eqnarray} && 
\qquad \qquad D^{(2)}
= (M D^{(0)}/4) < \bar X_{3}^{(1)} x_{1} \bar X_{4}^{(1)} x_{2} > \ ,
\nonumber \\ && 
D^{(4)}
= (M^{2}D^{(0)}/16) < X_{3}^{(1)} \bar x_{2} X_{4}^{(1)} \bar x_{1} >
  < \bar X_{3}^{(1)} x_{1} \bar X_{4}^{(1)} x_{2} > \ , \qquad
\nonumber
\label{D2-D4/t}
\end{eqnarray}
where $ X_{s}^{(1)}, \bar X_{s}^{(1)} $ are the matrices given in   
Eq.~(\ref{Xs1-bXs1}) and $ x_{r}, \bar x_{r}, y_{r} $ the matrices
introduced in Eq.~(\ref{xq-bxq-yq}). For brevity, we present here only
those parts of $ D^{(2)}, D^{(4)} $ which give a nonvanishing
contribution to the integral.

For the boundary terms, inserting the matrices $V^{(n)}$ and applying
the projectors $\prod \nolimits_{r}^{(1)}P(z_{r})$ yields
\begin{eqnarray} &&
{\cal I }_{1011}^{(4)} [ F ]  
= (M^{2}/256) \int \mbox{d}[z_{1}] \prod \nolimits_{s} \mbox{d}[z_{s}]
\rho_{1011}^{(0)} K^{(0)} D_{0}
\prod \nolimits_{p} \gamma^{p}_{1} \gamma^{p\ast}_{1} \qquad \qquad 
\nonumber \\ && \qquad 
\times \ ( \ -2 < X_{3}^{(1)} k \bar X_{3}^{(1)} y_{1} >
  < X_{4}^{(1)} y_{1} \bar X_{4}^{(1)} k > 
\nonumber \\ && \qquad \qquad \quad
+ \ M^{2} < y_{1} >
  < X_{3}^{(1)} k \bar X_{3}^{(1)} x_{1}
  \bar X_{4}^{(1)} k X_{4}^{(1)} \bar x_{1} >
\nonumber \\ && \qquad \qquad \qquad \quad
+ \ \ 2 M < X_{3}^{(1)} k \bar X_{3}^{(1)} x_{1}
  \bar X_{4}^{(1)} k X_{4}^{(1)} y_{1} \bar x_{1} >
\quad ) \ ,
\nonumber \\ &&
{\cal I }_{0011}^{(4)} [ F ]
= ( M^{2}/512 ) \int \prod \nolimits_{s} \mbox{d} [ z_{s} ] 
\rho_{0011}^{(0)} K^{(0)} D_{0} 
\nonumber \\ && \qquad \qquad
\times < X_{3}^{(1)} k \bar X_{3}^{(1)} k >
< X_{4}^{(1)} k \bar X_{4}^{(1)} k > \ ,
\label{IanF-Xnxn/bout}
\end{eqnarray}
with $\rho_{a}^{(0)} = \prod \nolimits_{r}^{(1)} 
\rho_{\mbox{\scriptsize G}}( \theta_{r} ) 
\prod_{s} \rho_{\mbox{\scriptsize G}}^{(0)}( \theta_{s} )$ . 
The phases and the anticommuting variables can be integrated out 
using the integration formulae~(\ref{int-ch}). This simplifies the
boundary terms to the eigenvalue integrals
\begin{eqnarray} &&
{\cal I }_{1011}^{(4)} [ F ]
= ( M^{2}/64 ) \int \mbox{d}[ \theta_{1} ]
\prod \nolimits_{s} \mbox{d}[ \theta_{s} ]
\ \rho_{1011}^{(0)}
K^{(0)} D_{0} \prod \nolimits_{s} < (\theta_{s})^{2} > \qquad \qquad
\nonumber \\ && \qquad \qquad \qquad
\times \Big( - (M^{2} + 2) < y_{1} >^{2} 
+ 2M < y_{1} ( 1 - y_{1} ) > \Big) \ ,
\nonumber \\ &&
{\cal I }_{0011}^{(4)} [ F ] 
= (M^{2}/32)
\int \prod \nolimits_{s} \mbox{d}[ \theta_{s} ] \rho_{0011}^{(0)}
K^{(0)} \prod \nolimits_{s} < (\theta_{s})^{2} > \ . 
\label{IanF-th/bout}
\end{eqnarray}
These integrals can be done using Eqs.~(\ref{int-thr/t}) and
(\ref{int-ths/t}).   

\subsection{The Limit of Large $M$ at Fixed $t/M$}
\label{M>>1}

This limit has been considered, and the leading term worked out,
previously in Refs.~\cite{efe95,fra95}. Thus, the present Appendix
provides a test of our procedure. At the same time, comparison between
the two approaches shows the relative simplicity of the present
approach. We proceed in the same way as in Section \ref{t>>1}. 
Many steps are similar and will not be repeated. We expand the
autocorrelation function $C(t,M)$ in inverse powers of $M$ and
evaluate the leading term. In the limit where $M,t \gg 1$ at $t/M$
fixed and arbitrary, the integrals ${\cal I}_{a}[F]$ appearing in the 
sum~(\ref{deg12-IaF}), i.e. the integrals with $ a = (1111), (1011),
(0011) $, are dominated by the contribution of the neighbourhood of
the surface $\theta_{r}^{\alpha} = \theta_{s}^{\alpha} = 0$ where
$Q=L$ and where the graded traces in the exponents of
\begin{equation} 
K= \mbox{e}^{ -(t/2)<(Q\tau_{3})^{2}> } \ , \qquad
D= \mbox{e}^{ - M <\ln(1+QL)> } 
\label{KD-Q}
\end{equation}
are equal to zero.

\subsubsection{ Asymptotic Expansion}
\label{asy-exp/M}

We start from Eq.~(\ref{IaF-PKSD}), rescale the angles $ \tilde
\theta_{q}^{\alpha} = M^{1/2} \theta_{q}^{\alpha} $, and pass from the
coordinates $ z_{q} $ to the coordinates $ \tilde z_{q} = ( \tilde
\theta_{q}^{b}, \phi_{q}^{b}, \tilde \theta_{q}^{f}, \phi_{q}^{f},
\gamma_{q}^{1}, \gamma_{q}^{1\ast}, \gamma_{q}^{2}, \gamma_{q}^{2\ast}
)$ , with $\mbox{d}[z] \rho_{a} \prod \nolimits_{r}^{(0)}\delta(z_{r}) =
\mbox{d}[\tilde z] M^{-l}\rho_{a} \prod \nolimits_{r}^{(0)}\delta(\tilde
z_{r})$ for $ l = \sum_{q} a_{q} $. At fixed $t/M$, we expand $ M^{-l}
\rho_{a} $, $ K $, $ V $ and $ D $ in powers of $ M^{-1/2} $.
The expansion coefficients $\rho_{a}^{ (n_{\rho}) },K^{(n_{k})},
V^{(n_{rp})},D^{(n_{d})}$ are functions of $\tilde
z_{q}$. We take into account that the $N_{j}$ also depend on $M$, and
write $ N_{j} = M^{ \nu_{j} /2 } N_{j}^{ ( \nu_{j} ) } $, with $
\nu_{j} = 4, 8, 6, 6 $ for $ j = 1,2,3,4 $, respectively. We collect
all terms  of the same order in $ M^{-1/2} $ and get
\begin{equation}
{\cal I}_{a}[ F ] = \sum_{n} M^{-n/2} {\cal I}_{a}^{(n)}[ F ] \ .
\label{IaF-IanF/M}
\end{equation}
Here ${\cal I}_{a}^{(n)}[ F ]$ denotes the sum of integrals
\begin{equation}
{\cal I}_{a}^{(n)}[ F ]
= \sum_{ j} \sum_{ n_{\rho} n_{k} n_{11} n_{12} n_{21} n_{22} n_{d} }
{\cal I}_{a}^{ ( n_{\rho}; n_{k}; n_{v}; n_{d}: \nu_{j}) }[ F_{j} ] \
\label{IanF-IanFj/M}
\end{equation}
with
\begin{eqnarray} &&
{\cal I}_{a}^{ ( n_{\rho}; n_{k}; n_{v}; n_{d}: \nu_{j} ) }[ F_{j} ]
= \int \mbox{d}[ \tilde z ] \rho_{a}^{ ( n_{\rho} ) }
\nonumber \\ && \qquad\times
\prod \nolimits_{r}^{(0)} \delta ( \tilde z_{r} ) \ 
\prod \nolimits_{r}^{(1)} P ( \tilde z_{r} ) \
\prod \nolimits_{s} P ( \tilde z_{s} ) \
K^{ ( n_{k} ) } N_{j}^{(\nu_{j})} S_{j}^{(n_{v})} D^{ ( n_{d} ) } \ . \qquad
\label{IanFj/M}
\end{eqnarray}
For fixed $j$, the sum over $ n_{\rho}, n_{k}, n_{rp}, n_{d} $ is
restricted by the condition $ n_{\rho} + n_{k} + n_{\sigma}+ n_{d} = n
+ \nu_{j}$. All integrals contain the exponential factor
$K^{(0)}D^{(0)}$ with 
\begin{equation}
K^{(0)} =
\mbox{e}^{ (2t/M)\sum_{s} < ( \tilde \theta_{s} )^{2} > } \ ,
\qquad
D^{(0)} =
\mbox{e}^{ (1/4)\sum_{q} < ( \tilde \theta_{q} )^{2} > } \ .
\label{K0D0/M}
\end{equation}
On extending the domain of integration region over $\tilde
\theta_{q}^{f}$ from zero to infinity, the series~(\ref{IaF-IanF/M})
yields an asymptotic expansion of ${\cal I}_{a}[ F ]$ .
Since $\rho_{a}$, $K$, $S$ and $D$ are even functions of
$\theta_{q}^{\alpha}$, only the terms with even $n$ appear, and the
expansion proceeds in inverse powers of $M$,
\begin{equation}
C(t,M) 
= \sum_{n=1}^{\infty} M^{-n} \bigg(\ {\cal I}_{1111}^{(2n)}[ F ]
   + 2 {\cal I}_{1011}^{(2n)}[ F ]
   + {\cal I}_{0011}^{(2n)}[ F ] \  \bigg) \;.
\label{deg12-asyexp/M}
\end{equation}
The integrals $ {\cal I}_{a}^{ ( n_{\rho}; n_{k}; n_{v}; n_{d}: n_{j}
  ) }[ F_{j} ] $ satisfy the same symmetry relations as the integrals
$ {\cal I}_{a}^{ ( n_{\rho}; n_{k}; n_{v}; n_{d}) }[ F_{j} ] $,
cf. Eq.~(\ref{IanFj-sym/t}). The leading term of the expansion is of
the order  $ M^{0} $, and is given by the sum 
\begin{equation}
C_{ M \gg 1 }(t,M)
   = \ {\cal I}_{1111}^{(0)}[ F ]
   + 2 {\cal I}_{1011}^{(0)}[ F ]
   + {\cal I}_{0011}^{(0)}[ F ] \ .
\label{deg12-leater/M}
\end{equation}

\subsubsection{The Leading Term $ C_{ M \gg 1 }(t,M) $ }
\label{lea-ter/M}

The nonvanishing projections $ \sum_{s}P(z_{s}) S_{j}^{(n_{v})}
D^{(n_{d})} $ stem solely from the terms which are linear in all four
matrices $( \delta Q_{s}^{12} )^{(1)}$, $ ( \delta Q_{s}^{21} )^{(1)}
$ and which for each $r$ contain the same number of $ (
G_{r}^{12})^{(1)} $ and of $ ( G_{r}^{21})^{(1)}$. We employ the
symmetry properties of $ {\cal I}_{a}^{ ( n_{\rho}; n_{k}; n_{v};
  n_{d}; \nu_{j}) }[ F_{j} ] $ and find that 
\begin{eqnarray} &&
{ \cal I }_{1111}^{(0)} [ F ]
= { \cal I }_{1111}^{(0;0;1111;4:8)} [ F_{2} ]
\nonumber \\ && \quad
+\, 4 { \cal I }_{1111}^{(0;0;1113;2:8)} [ F_{2} ]
+ 2 { \cal I }_{1111}^{(0;0;1331;0:8)} [ F_{2} ]
+ 2 { \cal I }_{1111}^{(0;0;1111;2:6)} [ F_{3} ] \ , \qquad
\label{IanF-IanFj-sym/volM}
\end{eqnarray}
\begin{eqnarray} &&
{ \cal I }_{1011}^{(0)} [ F ] 
= { \cal I }_{1011}^{(0;0;1133;0:8)} [ F_{2} ]
+ 2 { \cal I }_{1011}^{(0;0;1131;0:6)} [ F_{3} ] \ ,
\nonumber \\ &&
{ \cal I }_{0011}^{(0)} [ F ] =
{ \cal I }_{0011}^{(0;0;1111;0:4)} [ F_{1} ] \ .
\label{IanF-IanFj-sym/bouM}
\end{eqnarray}
For the volume term ${\cal I}_{1111}^{(0)}[F]$,
\begin{displaymath}
V^{(1)} = \frac{1}{4}
     \left(
          \begin{array}{cccc}
          0 & 2 x_{1}^{(1)} & 0 & X_{3}^{(1)} \\
          2 \bar x_{1}^{(1)} & 0 & - \bar X_{4}^{(1)} & 0 \\
          0 & X_{4}^{(1)} & 0 & 2 x_{2}^{(1)} \\
          - \bar X_{3}^{(1)} & 0 & 2 \bar x_{2}^{(1)} & 0
          \end{array}
     \right) \ ,
\end{displaymath}
\begin{displaymath}
V^{(3)} = - \frac{1}{8}
     \left(
          \begin{array}{cccc}
          0 & X_{3}^{(1)} \bar x_{2}^{(1)} X_{4}^{(1)} & 0 & 0 \\
          \bar X_{4}^{(1)} x_{2}^{(1)} \bar X_{3}^{(1)}  & 0 & 0 & 0 \\
          0 & 0 & 0 & X_{4}^{(1)} \bar x_{1}^{(1)} X_{3}^{(1)}  \\
          0 & 0 & \bar X_{3}^{(1)} x_{1}^{(1)} \bar X_{4}^{(1)} & 0
          \end{array}
     \right) \ ,
\end{displaymath}
\begin{eqnarray} && \qquad \qquad
D^{(2)}
= (D^{(0)}/4) < \bar X_{3}^{(1)} x_{1}^{(1)}
\bar X_{4}^{(1)} x_{2}^{(1)} > \ ,
\nonumber \\ &&
D^{(4)}
= (D^{(0)}/16) < X_{3}^{(1)} \bar x_{2}^{(1)} X_{4}^{(1)} \bar x_{1}^{(1)} >
  < \bar X_{3}^{(1)} x_{1}^{(1)} \bar X_{4}^{(1)} x_{2}^{(1)} > \ , \qquad
\nonumber
\end{eqnarray}
where $ X_{s}^{(1)}, \bar X_{s}^{(1)} $ are the matrices introduced
in Eq.~(\ref{Xs1-bXs1}), and $ x_{r}^{(1)}, \bar x_{r}^{(1)} $
denote the matrices
\begin{equation}
x_{r}^{(1)} = (G_{r}^{12})^{(1)} =
i (\theta_{r}/2) \mbox{e}^{i\phi_{r}}, \qquad
\bar x_{r}^{(1)} = (G_{r}^{21})^{(1)} =
i (\theta_{r}/2) \mbox{e}^{-i\phi_{r}} \ .
\label{xr1-bxr1}
\end{equation}
We present only those parts of $V^{(3)},D^{(2)},D^{(4)}$ which give
nonzero contribution to the integrals. The densities $\rho_{a}^{(0)}$
have for all $a$ the form $\rho_{a}^{(0)} = \prod \nolimits_{q}^{(1)} 
\rho_{\mbox{\scriptsize G}}^{(0)}( \theta_{q} )$, with
$\rho_{\mbox{\scriptsize G}}^{(0)}$ given in Eq.~(\ref{rh0/volt}). The
integrals are evaluated in the same way as in Subsection \ref{lea-ter/t}.
The integration over the eigenvalues is done making use of
\begin{eqnarray} &&
\int \mbox{d} [ \theta_{r} ]
\rho_{ \mbox{ \scriptsize G } }^{(0)} ( \theta_{r} )
\mbox{e}^{\frac{1}{4} < (\theta_{r})^{2} > }
< y_{r}^{(2)} > \  = 1 \ ,
\nonumber\\ &&
\int \mbox{d} [ \theta_{r} ]
\rho_{ \mbox{ \scriptsize G } }^{(0)} ( \theta_{r} )
\mbox{e}^{\frac{1}{4} < (\theta_{r})^{2} > }
< y_{r}^{(2)} >^{2} \ = - 1  
\label{int-thr/M}
\end{eqnarray}
and
\begin{equation}
\int \mbox{d} [ \theta_{s} ]
\rho_{ \mbox{ \scriptsize G } }^{(0)} ( \theta_{s} )
\mbox{e}^{\frac{1}{4}[1 + (8t/M)] < (\theta_{s})^{2} > }
< (\theta_{s})^{2} > \
= 4 \Big[ 1 + ( 8t/M ) \Big]^{-1} \ . \qquad
\label{int-ths/M}
\end{equation}
The contributions stemming from the
integrals appearing in the second row of 
Eq. (\ref{IanF-IanFj-sym/volM}) cancel mutually, 
as well as the contributions
of the two boundary terms. 
Thus the evaluation of the leading term reduces to 
the evaluation of the integral
\begin{eqnarray} &&
{\cal I}_{1111}^{(0;0;1111;4;8)}[F_{2}] =
\int \mbox{d}[ z ] \rho_{1111}^{(0)} K^{(0)}  
\nonumber \\ && \qquad
\times 
\ N_{2}^{(8)} < V^{(1)}\, I_{0}^{(11)}\, V^{(1)}\, I_{0}^{(12)} >
      < V^{(1)}\, I_{0}^{(21)}\, V^{(1)}\, I_{0}^{(22)} > D^{(4)}
\nonumber \\ && \qquad 
= (1/64) \int \mbox{d}[z] \rho_{1111}^{(0)}  
K^{(0)} D^{(0)} \prod \nolimits_{r}< y_{r}^{(2)} >^{2}
\nonumber \\ && \qquad
\times \ \prod \nolimits_{ rp } \gamma^{p}_{r} \gamma^{p\ast}_{r}
< X_{3}^{(1)} \bar x_{2}^{(1)} X_{4}^{(1)} \bar x_{1}^{(1)} >
< \bar X_{3}^{(1)} x_{1}^{(1)} \bar X_{4}^{(1)} x_{2}^{(1)} >  \qquad
\nonumber \\ && \qquad 
= (1/64)\! \int \! \prod \nolimits_{q} \mbox{d}[ \theta_{q} ] \rho_{1111}^{(0)}
K^{(0)} D^{(0)} \prod \nolimits_{r} \! < y_{r}^{(2)} >^{2}
\prod \nolimits_{s} \! < (\theta_{s})^{2} > \ .\qquad
\label{lea-ter-int/M}
\end{eqnarray}
The integrals over $\theta_{r}$ yield the unit diffuson factor
independent of $t/M$, the integrals over $\theta_{s}$ yield the
cooperon factor $16\Big( 1+(8t/M) \Big)^{-2}$.
The result is
\begin{equation}
C_{M>>1}(t,M) = \frac{1}{4}
\Bigg( \ 1 + \frac{8t}{M} \ \Bigg)^{-2} \ .
\label{IanF-res/volM}
\end{equation}
This result agrees with the one obtained by Efetov \cite{efe95}
and by Frahm \cite{fra95} in the framework of supersymmetry, and with
the squared Lorentzian found in the semiclassical approach
\cite{jal90}.

\end{document}